\documentclass[prd,aps,showpacs,tightenlines, 10pt, reprint,nofootinbib,superscriptaddress,floatfix,twocolumn]{revtex4-2}

\usepackage[latin1]{inputenc} 

\usepackage{acronym}
\usepackage{amsmath}
\usepackage{amsthm}  
\usepackage{amssymb}
\usepackage{mathtools} 
\usepackage{phaistos}
\usepackage{graphicx}
\usepackage{subcaption}
\usepackage[section]{placeins}

\usepackage{physics} 
\usepackage{IEEEtrantools}  
\usepackage{xspace}
\usepackage{cancel}
\usepackage[normalem]{ulem} 

\usepackage{epsfig}
\usepackage{multirow} 
\usepackage{color}
\usepackage[dvipsnames]{xcolor}
\usepackage{rotating}
\usepackage[export]{adjustbox} 
\usepackage{floatrow}
\newfloatcommand{capbtabbox}{table}[][\FBwidth]
\usepackage{tabularx}
\usepackage[pdfencoding=auto, psdextra]{hyperref}
\usepackage{units}
\usepackage[T1]{fontenc}
\usepackage[normalem]{ulem} 
\usepackage{aas_macros}

\newcommand{\nn}{\nonumber}

\newcommand{\dm}{\text{dm}}
\newcommand{\typeI}{\text{Type--I}}
\newcommand{\ostriker}{\text{Ostriker}}
\newcommand{\gw}{\text{gw}}
\newcommand{\orb}{\text{orb}}
\renewcommand{\b}{\text{b}}
\newcommand{\isco}{\text{isco}}
\renewcommand{\sp}{\text{spike}}
\newcommand{\edd}{\text{Edd}}
\newcommand{\vac}{\text{vacuum}}
\newcommand{\E}{\mathcal{E}}
\newcommand{\F}{\mathcal{F}}

\DeclarePairedDelimiterX{\avg}[1]{\langle}{\rangle}{#1}
\makeatletter
\let\oldavg\avg
\def\avg{\@ifstar{\oldavg}{\oldavg*}}  

\renewcommand{\eqref}[1]{Eq.~(\ref{#1})}
\newcommand{\neqref}[1]{(\ref{#1})}
\newcommand{\figref}[1]{Fig.~\ref{#1}}
\newcommand{\lcdm}{$\Lambda$CDM}

\begin{document}

\title{Comparing Accretion Disks and Dark Matter Spikes in Intermediate Mass Ratio Inspirals}
\author{Niklas Becker}
\email{nbecker@itp.uni-frankfurt.de} 
\affiliation{Institute for Theoretical Physics, Goethe University, 60438 Frankfurt am Main, Germany}

\author{Laura Sagunski}
\email{sagunski@itp.uni-frankfurt.de} 
\affiliation{Institute for Theoretical Physics, Goethe University, 60438 Frankfurt am Main, Germany}

\date{\today}

\begin{abstract}
Intermediate Mass Ratio Inspirals (IMRIs) will be observable with space-based gravitational wave detectors such as the Laser Interferometer Space Antenna (LISA). To this end, the environmental effects in such systems have to be modeled and understood. These effects can include (baryonic) accretion disks and dark matter (DM) overdensities, so called \textit{spikes}. For the first time, we model an IMRI system with both an accretion disk and a DM spike present and compare their effects on the inspiral and the emitted gravitational wave signal. We study the eccentricity evolution, employ the braking index and derive the dephasing index, which turn out to be complementary observational signatures. They allow us to disentangle the accretion disk and DM spike effects in the IMRI system and can be utilized to study environmental effects in general. 
\end{abstract}
\preprint{}
\maketitle

\section{Introduction \label{sec:intro}} 
The first direct detection of gravitational waves (GWs) has opened a fundamentally new window into the Universe. The Laser Interferometer Gravitational-Wave Observatory (LIGO) has seen the first binary black hole merger, and, together with the LIGO-Virgo-KAGRA (LVK) collaboration, has already collected a sizable catalogue of compact binary mergers \cite{LIGOScientific:2016aoc, LIGOScientific:2021djp}. They allow new and unprecedented tests of General Relativity and matter at high densities \cite{LIGOScientific:2020tif, LIGOScientific:2018cki}. On top of that, there are several space-based observatories planned such as LISA \cite{LISA:2017pwj}, Taiji \cite{10.1093/nsr/nwx116} and TianQuin~\cite{TianQin:2015yph}, which will allow for the detection of GWs at lower frequencies. For these observatories, accurate waveforms have to be computed to maximize the science yield \cite{Zwick:2022dih}.

While LVK mostly observes solar mass binary mergers, space-based observatories will be able to detect Extreme/Intermediate mass ratio inspirals (E/IMRIs). In these systems, a stellar mass object inspirals into a supermassive/intermediate mass black hole (S/IMBH). Several IMBH candidates have been detected, but their origin, evolution, and environment is not yet well understood \cite{Lin:2018dev, Mezcua:2017npy}. To observe these IMRI systems, the environmental effects have to be understood first, as accurate waveforms are needed for detection \cite{Zwick:2022dih}.

Meanwhile, dark matter (DM) as predicted by \lcdm{} has continued to elude detection \cite{Bertone:2004pz, XENON:2018voc}. While its effects are observed on large scales, such as structure formation, on small scales, the effects of dark matter are more uncertain and a plethora of models has been proposed \cite{Bertone:2018krk}. Around IMBHs, on small scales, a dark matter halo could grow adiabatically into a dark matter \textit{spike} \cite{Gondolo:1999ef, Sadeghian:2013laa}. These spikes have an extremely high local density compared to the ambient DM density and would gravitationally interact with any object passing through. During an IMRI, the dark matter spike leaves its imprint by modifying the orbital evolution. This is one of several possible environmental effects.

This has first been explored in \cite{Eda:2013gg, Eda:2014kra}, where the authors predicted a dephasing of the GW signal due to dynamical friction with the DM spike \cite{Chandrasekhar:1943ys}. If the secondary object is a black hole, it will accrete the DM as it passes through the spike, which was first explored in \cite{Macedo:2013qea} and later in \cite{Yue:2017iwc}, where the accretion effects were found to be subdominant to dynamical friction effects. Then, \cite{Yue:2019ozq, Cardoso:2020iji} looked at eccentric orbits, and found there to be an eccentrification of the orbits. Afterwards, we have argued in \cite{Becker:2021ivq} that by including the phase space distribution of dark matter particles, the system circularizes. Other spike effects have been studied, such as periastron precession\cite{Dai:2021olt, Destounis:2022obl}, halo feedback mechanism \cite{Kavanagh:2020cfn, Coogan:2021uqv}, relativistic corrections to dynamical friction and spike distribution \cite{Speeney:2022ryg}, and spikes around lower mass primordial black holes \cite{Cole:2022ucw}.

Another source of important environmental effects is the presence of (baryonic) accretion disks \cite{Barausse:2007dy, Barausse:2014tra}. While the existence of DM spikes around IMBH is still speculative, the existence of accretion disks around SMBHs is supported observationally \cite{Padovani:2017zpf, EventHorizonTelescope:2019dse}, and a strong argument can be made for their existence around IMBHs as well. The effects of the interaction between the secondary object and the accretion disk in an IMRI can also affect the inspiral. While interesting from a physical standpoint, from the perspective of trying to detect dark matter, these baryonic effects could mimick or dominate dark matter effects, spoiling its detection. There have been studies trying to map out disk effects in IMRIs\cite{Derdzinski:2020wlw, Speri:2022upm, Cole:2022fir} and in this paper, for the first time, we want to compare the environmental effects of accretion disks and DM spikes. This allows us to estimate their relative strength and observational signatures.


The motivation of this paper is to model IMRIs on eccentric Keplerian orbits with GW emission, dynamical friction with the DM spike, and gas interaction with the accretion disk. We do not include all relevant effects, such as halo feedback and relativistic corrections here, but focus on comparing baryonic and dark matter effects first. More expansive studies are left for future work.

The structure of the paper is as follows. In section \ref{sec:equations}, we explain the theoretical framework to model the orbital evolution of the IMRI and its GW signal. In section \ref{sec:results}, we present numerical results, and discuss them in section \ref{sec:discussion}. Finally, conclusions are drawn in section \ref{sec:concl}. 

Throughout the paper we adopt geometrized units with $G = c = 1$.


\section{IMRI Modeling \label{sec:equations}}

\begin{figure}
    \centering
    \includegraphics[width=0.9\textwidth]{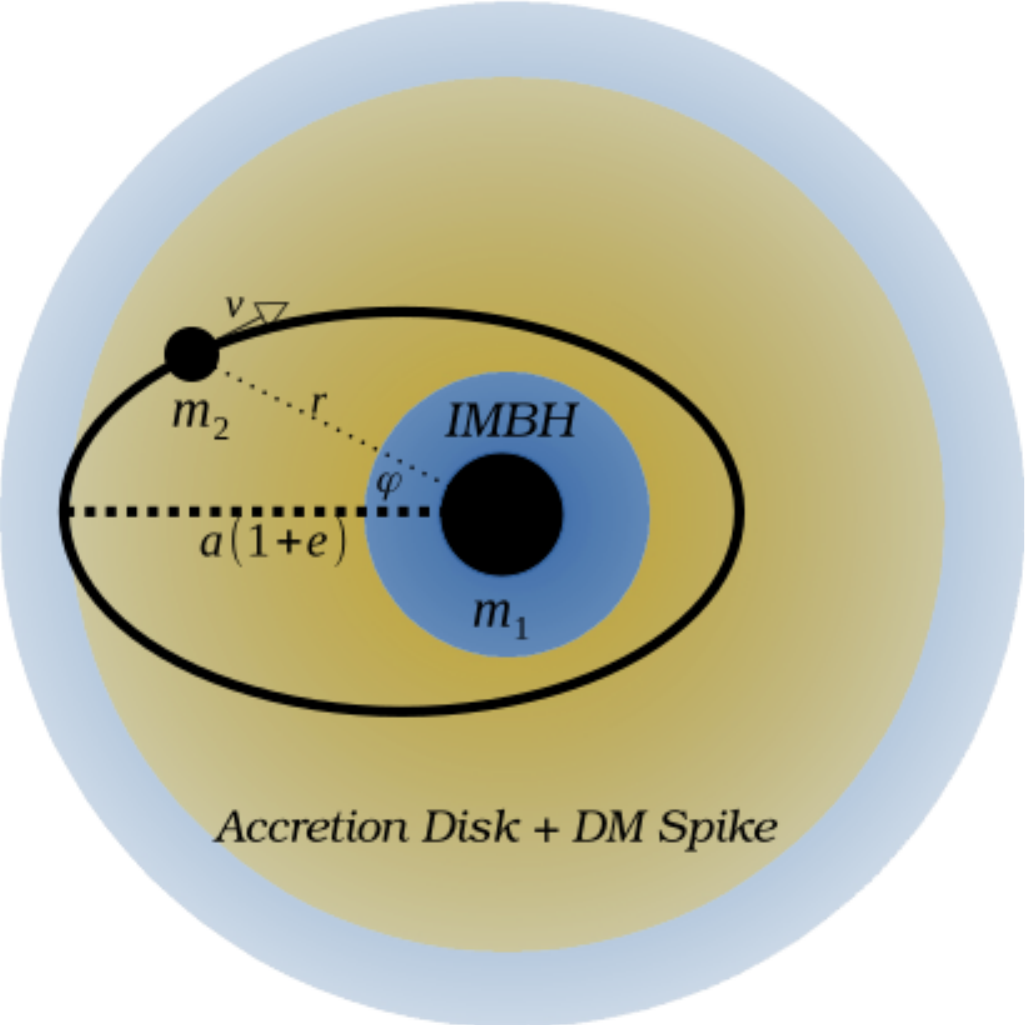}
    \caption{A sketch of the IMRI system with masses $m_1 \gg m_2$, semimajor axis $a$, and eccentricity $e$ inside the DM spike $\rho_\dm$ and accretion disk $\rho_\b$. Here, the true anomaly $\phi = \pi - \varphi$, such that $\phi = 0$ is the pericenter, and $\phi = \pi$ the apocenter. }
    \label{fig:sketch}
\end{figure}

The IMRI system consists of a central IMBH $m_1$ and a secondary object $m_2$, both of which are assumed to be Schwarzschild black holes. See \figref{fig:sketch} for a sketch. The IMBH is surrounded by both an axially symmetric accretion disk, and a spherically symmetric DM spike. The secondary is assumed to be on a Keplerian orbit around the central mass. The system emits GWs and is subject to environmental effects, such as those given by the interactions with the DM spike and accretion disk. Through these dissipative forces, the secondary loses orbital energy and angular momentum, leading to an inspiral. In this section we present the theoretical background and observational signatures.

\subsection{Dark Matter Spike}
To model the DM spike, we follow our previous publication \cite{Becker:2021ivq}. The development and existence of DM spikes has been discussed extensively in the literature \cite{Gondolo:1999ef, Ullio:2001fb, Sadeghian:2013laa, Coogan:2021uqv}.

We consider a system in which the IMBH $m_1$ is assumed to be surrounded by a static, spherically symmetric DM spike. We describe the spike density around the central mass by a simple power law \cite{Coogan:2021uqv}
\begin{equation}
\label{eq:rho_dm}
    \rho_\dm(r) = \rho_6 \left(\frac{r_6}{r}\right)^{\alpha_\sp}, \quad  r_\text{in} < r < r_\sp 
\end{equation}
with the radius from the central mass $r$ and the reference radius $r_6 = 10^{-6}$pc. Following \cite{Sadeghian:2013laa}, the inner radius is chosen to be $r_\text{in}=4m_1$. The spike radius $r_\sp $ is the maximal radius of the spike, which can be obtained by comparing the gravitational influence of the IMBH to the total spike mass \cite{Eda:2014kra}. In this publication, we always consider $r \ll r_\sp $. 

The range of the power law index is $1 < \alpha_\sp < 3$, but we focus on the $\alpha_\sp=7/3$ case in this paper, which represents a halo grown from an Navarro-Frenk-White (NFW) profile \cite{Navarro:1995iw}. See \cite{Eda:2014kra, Becker:2021ivq} for an exploration of different power laws.

The dark matter particles in the halo can be described by an equilibrium phase space distribution function $f=dN/d^3rd^3v$, giving the number density per phase space volume. Since the halo is spherically symmetric,  $f = f(\E)$, with $\E$ being the relative energy per unit mass
\begin{equation}
    \E(r,v) = \Psi(r) - \frac{1}{2}v^2.
\end{equation}
$\Psi(r)$ is the relative Newtonian potential. Close to the IMBH, it is simply $\Psi(r) = \frac{m_1}{r}$. For gravitationally bound particles we have $\E > 0$. 

For a spherically symmetric density profile $\rho(r)$, the distribution function $f(\E)$ can be calculated by the Eddington inversion procedure \cite{1987gady.book}. For the power law spike, this gives
\begin{align}
\label{eq:f_powerlaw}
    f_\sp(\E) =& \frac{\alpha_\sp(\alpha_\sp-1)}{(2 \pi)^{3/2}} \rho_6 \left(\frac{r_6}{m_1}\right)^{\alpha_\sp} \nn \\
     & \times \frac{\Gamma(\alpha_\sp - 1)}{\Gamma(\alpha_\sp - \frac{1}{2})} \E^{\alpha_\sp-3/2}
\end{align}
where $\Gamma(x)$ is the Gamma function. 

The density for a given distribution function can be recovered through
\begin{equation}
\label{eq:rho_f}
    \rho(r) = 4\pi \int_0^{v_{\text{max}}(r)} v^2 f\left(\Psi(r) - \frac{1}{2}v^2 \right) dv,
\end{equation}
where the escape velocity is given by $v_{\text{max}}(r)=\sqrt{2\Psi(r)}$.

We ignore the relativistic effects on the dark matter spike, derived by \cite{Sadeghian:2013laa} and explored in \cite{Speeney:2022ryg} and leave them for future work.

\subsection{Accretion Disk}
To model the (baryonic) accretion disk, we follow the approach of \cite{Speri:2022upm}. We assume a radiatively efficient, geometrically thin accretion disk model, employing the so-called $\alpha$ and $\beta$ disk prescriptions derived by Shakura \& Sunyaev \cite{Shakura:1972te}, see \cite{Abramowicz:2011xu, Kocsis:2011dr} for a review. Here, radiatively efficient means that we assume the heat generated by viscosity at any given radius is immediately radiated away.

The disks are parameterized by the viscosity parameter $\alpha$, which is estimated to be around $\alpha \sim 0.01 - 0.1$ \cite{King:2007cu}. The surface density $\Sigma$ and scale height $H$ of the disk are then given by the parameterizations \cite{Speri:2022upm, 1981ApJ...247...19S}
\begin{align}
    \Sigma_\alpha \left[\frac{\text{kg}}{\text{m}^2}\right]=& 5.4 \cdot 10^{3} \left(\frac{\alpha}{0.1}\right)^{-1} \left(\frac{f_\edd}{0.1}\frac{0.1}{\epsilon}\right)^{-1} \left(\frac{r}{10 m_1}\right)^{3/2} \label{eq:alpha_disk}\\
    \Sigma_\beta \left[\frac{\text{kg}}{\text{m}^2}\right]=& 2.1 \cdot 10^{7} \left(\frac{\alpha}{0.1}\right)^{-4/5} \left(\frac{f_\edd}{0.1}\frac{0.1}{\epsilon}\right)^{3/5} \left(\frac{m_1}{10^6 M_\odot}\right)^{1/5}  \nonumber \\
    & \cross \left(\frac{r}{10 m_1}\right)^{-3/5}  \label{eq:beta_disk}\\ 
    H \left[M_\odot\right]= & 1.5 \left(\frac{f_\edd}{0.1}\frac{0.1}{\epsilon}\right) m_1, 
\end{align}
where $f_\edd$ is the fraction of the Eddington accretion rate that the central IMBH is accreting at, and $\epsilon$ describes the efficiency of mass-energy conversion into luminosity in the disk. We will assume $f_\edd = \epsilon = 0.1$ throughout this paper.

The corresponding disk density is $\rho_\b = \Sigma/2H$, and the Mach number $\mathcal{M}_a = r/H$. The simple scalings of the disk models are valid within approximately $r \leq 10^3 m_1$ \cite{Kocsis:2011dr}.

The models originate from the assumption that the central BH accretes at a steady rate $\dot{M}_\text{disk} = 3\pi \nu \Sigma$, where $\nu = \alpha c_s^2/\Omega$ is the kinematic viscosity, $c_s$ the sound speed, and $\Omega =\sqrt{m_1/r^3}$ the orbital frequency. The sound speed is in general given by $c_s^2 = \frac{p_\text{rad} + p_\text{gas}}{\rho}$ with the radiation pressure $p_\text{rad}$ and the thermal gas pressure $p_\text{gas}$. In $\beta$ disks, the thermal gas pressure is assumed to be dominant $p_\text{gas} \gg p_\text{rad}$. While analytic solutions to the $\alpha$ disk model have some thermal instabilities, it seems to be more physically realistic and a decent approximation in the radiation dominated regime \cite{Speri:2022upm}.

\subsection{Orbital Evolution \label{sec:orbital_ev}}

\subsubsection{Keplerian Orbit}
The secondary is assumed to be on a Keplerian orbit around the central IMBH. We ignore the additional matter contributions by the spike and disk to the total and reduced mass $\mu$ of the Keplerian system and assume $m=m_1 + m_2$, $\mu = \frac{m_1 m_2}{m}$. The mass ratio is defined as $q = \frac{m_2}{m_1}$. For the separations at play the system is clearly gravitationally dominated by the IMBH, and the total enclosed mass of the dark matter and baryon distributions up to the location of the secondary is much smaller than the mass of the central IMBH, $m_\dm(r=10^5r_{\isco}), m_\b(r=10^5r_{\isco}) \ll m_1$. Here, $r_\isco$ is the radius of the innermost stable circular orbit, which is $r_\isco=6m_1$ for a Schwarzschild black hole. According to \cite{Dai:2021olt}, the inclusion of the gravitational influence of the spike distribution would primarily lead to orbital precession, which we neglect in this paper. 

The Keplerian orbit can be described by two parameters, the semimajor axis $a$ and the eccentricity $e$. For a bound orbit, $0 \leq e < 1$, where $e=0$ describes a circular orbit.

The orbital energy is given by \cite{Maggiore:2007ulw}
\begin{equation}
    E_{\text{orb}} = -\frac{m\mu}{2a},
\label{eq:E_orbit}
\end{equation}
the angular momentum $L_{\text{orb}}$ by
\begin{equation}
    e^2 - 1 = \frac{2E_{\text{orb}}L_{\text{orb}}^2}{m^2 \mu^3}, 
\label{eq:L_orbit}
\end{equation}
and the mean orbital frequency by
\begin{equation}
\label{eq:Kepler_F}
        \mathcal{F} = \frac{1}{2\pi} \sqrt{\frac{m}{a^3}}.
\end{equation}
Throughout one orbit, the radius and the velocity of the secondary at the true anomaly $\phi$ is given by
\begin{align}
    r =& \frac{a(1-e^2)}{1+ e \cos{\phi}}, \label{eq:Kepler_r}\\
    v^2 =& m \left( \frac{2}{r} - \frac{1}{a} \right). \label{eq:Kepler_v}
\end{align}

\subsubsection{Dissipative Forces}
The secondary is assumed to lose energy on a secular timescale that is much larger than the orbital timescale. This assumption allows us to use the Keplerian orbits to calculate the dissipative forces acting on the secondary. Over many orbits, these lead to a change in the orbital parameters. 
To model the dissipative forces, we use the force $F(r, v)$ depending on the separation $r$ and the velocity $v$ of the secondary.

The energy and angular momentum loss for a given dissipative force can be obtained by averaging over one orbit with orbital period $T$ \cite{Yue:2019ozq},
\begin{align}
    \avg{\dv{E}{t}} =& \int_0^T \frac{\mathrm{d}t}{T} \dv{E}{t} =- \int_0^T \frac{\mathrm{d}t}{T} F(r,v)v,  \label{eq:avgdEdT} \\
    \avg{\dv{L}{t}} =& \int_0^T \frac{\mathrm{d}t}{T} \dv{L}{t} = - \sqrt{ma(1-e^2)}\int_0^T \frac{\mathrm{d}t}{T} \frac{F(r,v)}{v}. \label{eq:avgdLdT}
\end{align}
These integrals can be calculated with the help of Eqs.~\neqref{eq:Kepler_r} and \neqref{eq:Kepler_v} and
\begin{equation}
    \int_0^T \frac{\mathrm{d}t}{T} G(r(t),v(t)) = (1-e^2)^{\frac{3}{2}}\int_0^{2\pi} \frac{\mathrm{d}\phi}{2\pi}\frac{G(r(\phi), v(\phi))}{(1 + e \cos{\phi})^{2}},  
\end{equation} 
which is valid for an arbitrary function $G(r,v)$ \cite{Maggiore:2007ulw}.

Therefore, for a given dissipative force $F(r,v)$, we can compute the energy and angular momentum loss either analytically or numerically using Eqs. \neqref{eq:avgdEdT} and \neqref{eq:avgdLdT}. 

The specific dissipative effects considered here are GW emission, dynamical friction with the DM spike, and gas interaction with the accretion disk. Each can be modeled as a force which leads to a loss of orbital energy and angular momentum over secular timescales,
\begin{align}
    \dv{E_\orb}{t} = \avg{\dv{E_\gw}{t}} + \avg{\dv{E_\dm}{t}} +  \avg{\dv{E_\text{gas}}{t}}, \label{eq:dEorb_dt} \\
    \dv{L_\orb}{t} = \avg{\dv{L_\gw}{t}} + \avg{\dv{L_\dm}{t}} + \avg{\dv{L_{\text{gas}}}{t}}. \label{eq:dLorb_dt}
\end{align}
\\

\noindent\textbf{Gravitational Waves}
The GW emission loss is given by \cite{Maggiore:2007ulw}
\begin{align}
    \avg{\dv{E_\gw}{t}} =& - \frac{32}{5} \frac{\mu^2 m^3 }{a^5} \frac{1 + \frac{73}{24}e^2 +\frac{37}{96}e^4}{(1-e^2)^{7/2}}, \label{eq:dE_gw}\\
    \avg{\dv{L_\gw}{t}} =& - \frac{32}{5} \frac{\mu^2 m^{5/2} } {a^{7/2}} \frac{1 + \frac{7}{8}e^2}{(1-e^2)^2} .\label{eq:dLgw}
\end{align}
\\

\noindent\textbf{Dynamical Friction with the DM Spike}

The dynamical friction with the DM spike is given by the Chandrasekhar equation \cite{Chandrasekhar:1943ys, Kavanagh:2020cfn}
\begin{equation}
\label{eq:F_df}
    F_\dm(r, v) = 4\pi m_2^2 \rho_\dm(r) \xi(v)  \frac{\log \Lambda}{v^2}
\end{equation}
with the Coulomb logarithm $\log\Lambda$. Here, we adopt the value $\log \Lambda = \sqrt{\frac{m_1}{m_2}}$ \cite{Kavanagh:2020cfn}. The factor $\xi(v)$ accounts for the fact that the particles in the DM spike are moving with different velocities relative to the secondary\cite{1987gady.book, Kavanagh:2020cfn}, because physically, DM particles only scatter and absorb momentum from the secondary if they are moving with a velocity that is slower compared to it. 

To estimate the density of particles moving slower than the secondary travelling at $v$, we can use \eqref{eq:rho_f}
\begin{equation}
\label{eq:rho_xi}
    \rho_\dm(r)\xi(v) = 4\pi \int_0^{v} v'^2 f\left(\Psi(r) - \frac{1}{2}v'^2 \right) dv'.
\end{equation}

\noindent\textbf{Baryonic Disk Interaction}

There is a wide range of models for compact object (CO) -- accretion disk interactions. In this paper, we consider and compare two different models from different origins, \textit{Type--I} migration and \textit{dynamical friction} with the accretion disk. See section \ref{sec:typeIvsDF} for a discussion of applicability. Here, we primarily want to give the equations governing the two models.

The most commonly employed model originates from planetary formation models and is called \textit{Type--I} migration\cite{2002ApJ...565.1257T}. In planetary migration models, the important quantity is the torque acting on the secondary body.
The equation for this torque is given by \cite{2002ApJ...565.1257T}
\begin{equation}
    \Gamma_\typeI = \Sigma r^4 \Omega^2 q^2 \mathcal{M}_a^2 ,
\end{equation}
which can be translated into a force -- the language of our model -- by
\begin{equation}
    F_\typeI = \Gamma_\typeI q / r
\end{equation}
The derivation assumes the creation of density wave resonances in the disk by the secondary object, which causes a negative torque on the perturber, and thus an inspiral. Note that this equation is only valid for circular orbits. 

The second model we look at is that of \textit{dynamical friction} with the gas. 
To differentiate between the gas and DM dynamical friction, we call this model \textit{Ostriker}. The friction force is given by \cite{Ostriker:1998fa, 10.1093/mnras/stac1294}
\begin{equation}
    F_\ostriker =  4\pi m_2^2 \rho_\b(r)  \frac{I}{v_{rel}^2}
\end{equation}
with 
\begin{equation}
    I = \frac{1}{2}
    \begin{cases}
        \log\frac{1- v_{rel}/c_s}{1+v/c_s} - v_{rel}/c_s    & \textit{subsonic}\\
        \log (1-(v_{rel}/c_s)^{-2}) + \log \Lambda      & \textit{supersonic}
    \end{cases}
\end{equation}
Here, $v_{rel}$ refers to the relative velocity between the gas and the secondary object. For this model, we will assume the secondary object to be counter-rotating to the disk, to avoid the scenario of $v_{rel}=0$. This model allows for non-circular orbits.

The quantities $\dv{E_\text{gas}}{t}, \dv{L_\text{gas}}{t}$ in \eqref{eq:dEorb_dt},(\ref{eq:dLorb_dt}) then refer to either \textit{Type--I} or \textit{Ostriker} models, which will be made clear wherever relevant.

\subsubsection{Orbital Evolution }
We want to obtain the secular evolution of the orbital parameters $a(t),e(t)$ under the backreaction of our dissipative forces.

We use \eqref{eq:E_orbit} to obtain
\begin{align}
    \pdv{E_\orb}{a} =& \frac{m_2 m_1}{2a^2} \\
    \dv{a}{t} =& \dv{E_\orb}{t}   / \pdv{E_\orb}{a}. \label{eq:da_dt}
\end{align}
Similarly, the evolution for $e$ is derived from \eqref{eq:L_orbit} as
\begin{equation}
\label{eq:de_dt}
    \dv{e}{t} = -\frac{1-e^2}{2e} \left(\dv{E_{\text{orb}}}{t}/E_{\text{orb}} + 2\dv{L_{\text{orb}}}{t}/L_{\text{orb}} \right).
\end{equation}
Combining Eqs. \neqref{eq:da_dt} and \neqref{eq:de_dt} with Eqs. \neqref{eq:dEorb_dt} and \neqref{eq:dLorb_dt}, we have a system of differential equations that can be solved numerically.

\subsection{Gravitational Wave Signal}
The IMRI system emits GWs as a result of the change in the quadrupole moment. The equations governing the gravitational wave signal are given in \cite{Becker:2021ivq}. Here, we want to assume that we can measure a single IMRI system and resolve its frequency evolution $\F(t)$. We want to explore observational signatures and see if we can distinguish our models.

\subsubsection{Braking Index}
For large masses or large initial semimajor axes, the inspiral might take $t \gg 10$yrs, making it difficult to observe in its entirety. At the same time, at larger separations, GW emission might be subdominant to other dissipative losses, which can dictate the frequency evolution. A possible signature to consider is the evolution of the orbital frequency $\mathcal{F}$, which can be measured on shorter timescales \cite{Robson:2018ifk, Renzo:2021aho}. A useful quantity is the so called \textit{braking index}, which was first described in the context of neutron star spin-down \cite{Lu:2022oys} but can be applied to any signal with an evolution in time,
\begin{equation}\label{eq:braking_index}
    n_b = \frac{\mathcal{F} \ddot{\mathcal{F}}}{\dot{\mathcal{F}}^2}.
\end{equation}
As an example, when circular GW emission losses are dominant, $\dot{\mathcal{F}}\propto \mathcal{F}^{11/3}$ and therefore $n_b=11/3$ \cite{Cutler:1994ys}. 

Relating this to the semimajor axis with \eqref{eq:Kepler_F}, we have
\begin{equation}
    n_b = \frac{5}{3} - \frac{2}{3} \frac{a \ddot{a}}{\dot{a}^2}.
\end{equation}

Following the approach of \cite{Becker:2021ivq}, if we model a dissipative force as $F\sim r^{\gamma}v^{\delta}$, we have, to second order in eccentricity
\begin{align} \label{eq:brak_approx}
    \frac{a\ddot{a}}{\dot{a}^2} \approx& k_1 + 2ae\dv{e}{a} \left(\frac{1/2-k_1}{1 - e^2}  + \frac{k_2}{1+k_2 e^2} \right) \\
    \text{with } k_1 =& 2 + \gamma - \frac{\delta+1}{2}, \nonumber \\
    k_2 =& (3 + \gamma^2 + \gamma(3-2\delta) - 2\delta + \delta^2)/4 \nonumber
\end{align}
See appendix \ref{sec:app_deriv} for a more detailed computation.

Therefore, in the circular case ($e=0$), if a given dissipative force dominates the inspiral, the braking index will be this simple algebraic combination of parameters.
For eccentric inspirals, a measurement of the braking index and the eccentricity evolution $\dv{e}{a}$ allows to determine the dissipative force parameters $\gamma$ and $\delta$ that dominate the inspiral.

\subsubsection{Dephasing}
For smaller orbital separations, the IMRI is dominated by GW emission loss. To observe the effect that a subdominant dissipative force $F$ has on the evolution, we can look at the \textit{dephasing}. To this end, we compare the number of GW cycles completed in the cases with and without this force present, following \cite{Kavanagh:2020cfn}. We can do this for each harmonic individually between some initial time $t_\text{i}$ and final time $t_\text{f}$  with
\begin{equation}
    N(t_\text{f}, t_\text{i}) = 2 \int_{t_\text{i}}^{t_\text{f}} \F(t)\mathrm{d}t.
\end{equation}
Setting $t_\text{f} = t_\text{c}$ as the time of coalescence, we obtain
\begin{equation}
    \Delta N(t)= N_\vac(t_\text{c}, t) - N_\text{tot}(t_\text{c}, t).
\end{equation}
where $N_\vac$ is the phase accumulation where just GW emission is driving the inspiral, and $N_\text{tot}$ is the phase accumulation of a system where GW emission and another environmental effect -- that we will mark by the letter $F$ -- is present. The dissipative forces are additive, such that the frequency evolution can be written as $\dot{\F}_\text{tot} = \dot{\F}_\vac + \dot{\F}_F$. When the GW emission is dominant, we can write this such that $\dot{\F}_\text{tot} = \dot{\F}_\vac(1+\varepsilon)$, where $\varepsilon= \frac{\dot{\F}_F}{\dot{\F}_\vac}$. This gives us the second derivative of the dephasing
\begin{equation}
    \frac{1}{2}\dv[2]{\Delta N}{t} = \dot{\F}_\vac - \dot{\F}_\text{tot} = \varepsilon \dot{\F}_\vac .
\end{equation}

In the circular case, $\dot{\F}_\vac \sim \F_\vac^{11/3}$, and a calculation of $\varepsilon$ gives
\begin{equation}
    \varepsilon \propto \F_\vac^{-2 - 2k_1 /3}
\end{equation}
assuming our dissipative force has the form $F\sim r^{\gamma}v^{\delta}$ as in the previous section. See appendix \ref{sec:app_deriv} as well.

This results in the amount of dephasing being accumulated
\begin{equation} \label{eq:dN_df}
    \Delta N \propto \F_\vac^{(11-2k_1)/3}.
\end{equation}
Since $\varepsilon \ll 1$, $\F_\text{tot}(t) \approx \F_\vac(t)$, a measurement of 
\begin{equation}\label{eq:deph_index}
    n_d \equiv \dv{\log \Delta N}{\log \F_\text{tot}} = \F_\text{tot} \dv{\Delta N}{\F_\text{tot}}/\Delta N \approx \frac{11-2k_1}{3},
\end{equation}
what we will call the \textit{dephasing index}, could reveal the power law behavior of the dephasing that is accumulated, and therefore the dissipative force at play.

\section{Results \label{sec:results}}
In this section, we present the results from the numerical integration of the system of differential equations derived in sec. \ref{sec:orbital_ev}. The equations have been implemented in \textsc{python} and numerically solved. The code is publicly available at: \url{https://github.com/DMGW-Goethe/imripy}. 

\subsection{Accretion Disk Only}
\begin{figure}
    \centering
    \includegraphics[width=0.95\textwidth]{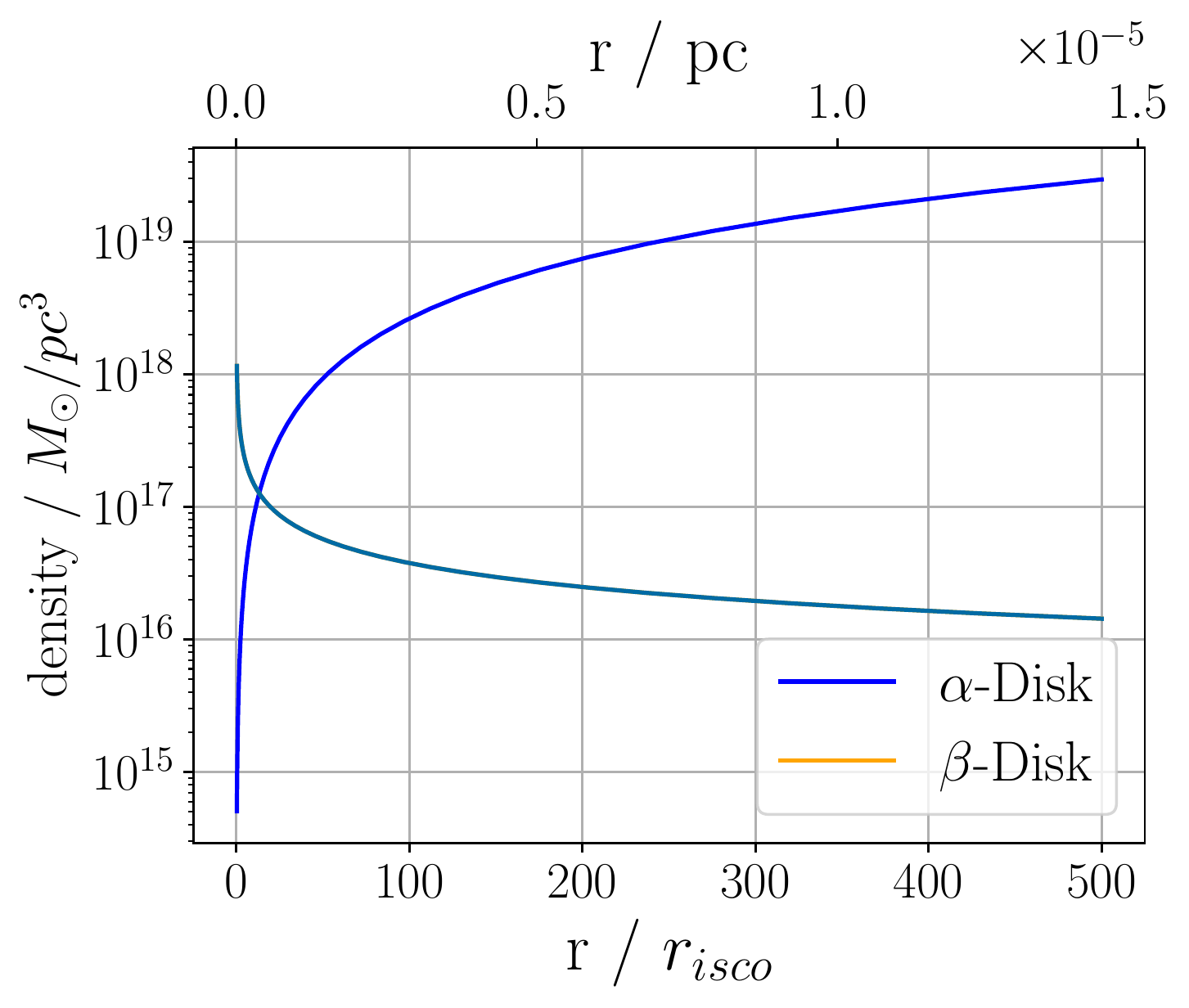}
    \includegraphics[width=0.95\textwidth]{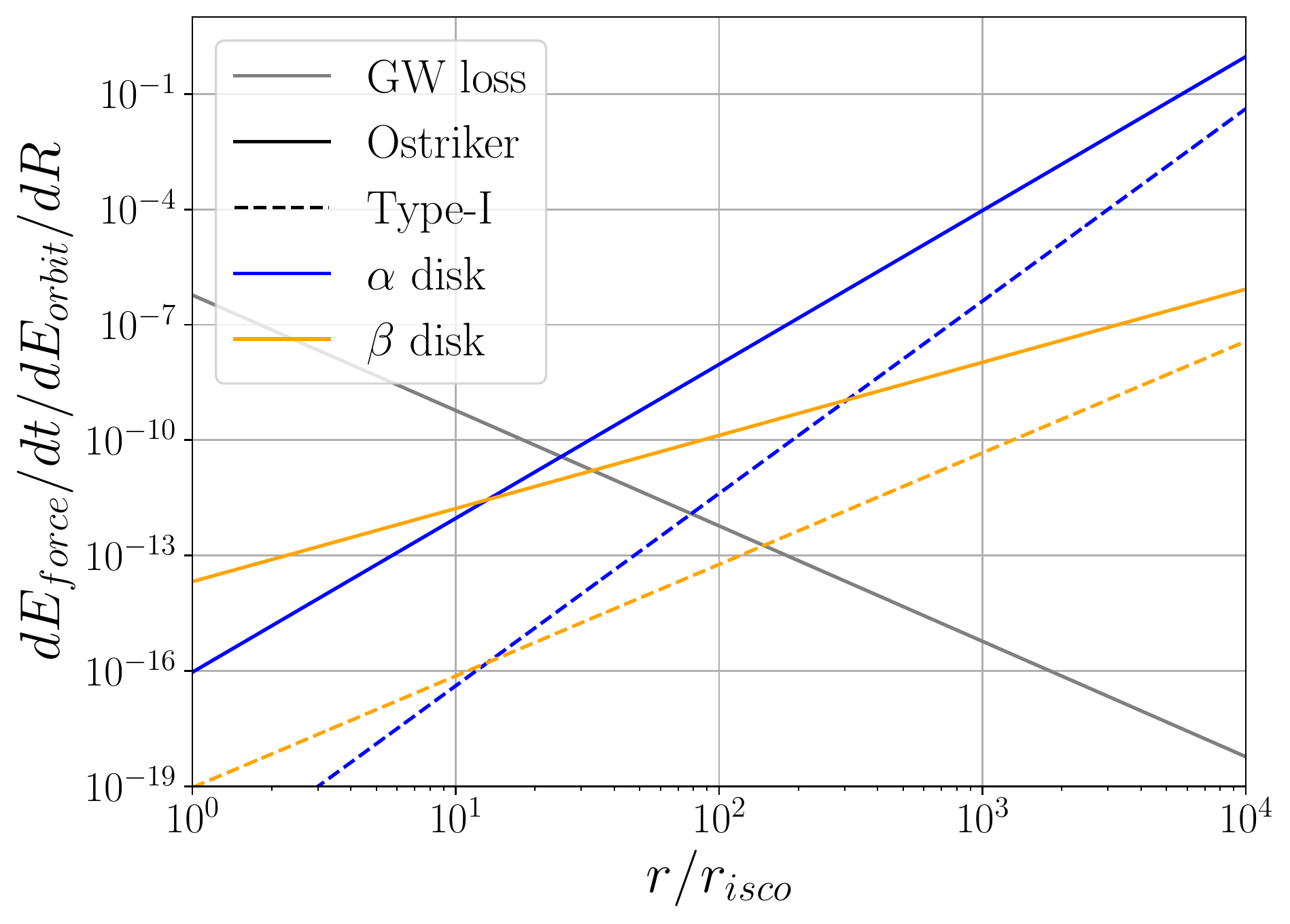}
    \caption{ \textbf{Top}: The density $\rho_\b$ of the two accretion disk models $\alpha$ and $\beta$ around an IMBH with $m_1=10^5 M_\odot$. \textbf{Bottom}: The relative impact of the three dissipative forces: GW emission loss, \textit{Type--I}, and \textit{Ostriker} for the two different disk models for a circular orbit at the given radius.  }
    \label{fig:avsb_T1vsOs}
\end{figure}

As a point of comparison, let us first consider an IMRI system with $\{m_1, m_2\} = \{10^5 M_\odot, 1 M_\odot\}$, in which the central mass is surrounded by an accretion disk. We model an $\alpha$ and a $\beta$-disk with the parameters $\alpha =0.1$. In \figref{fig:avsb_T1vsOs} we plot the density close to the $r_\isco$.

We first want to compare the dissipative force models \textit{Type--I} and \textit{Ostriker}. To get an estimate of the relative impact of the forces, we plot the relative semimajor axis loss for the forces involved $\dv{E_\text{force}}{t}/\pdv{E_\orb}{a}$, assuming circular orbits at a given radius. Close to $r_\isco$, the dominant force is the GW emission loss. The accretion disk effects become dominant for $r \gtrsim 10-100 r_\isco$. The steeper power law in the density of the $\alpha$ disk compared to the $\beta$ disk is reflected in the power law behavior of the energy loss curves. It can be seen that the \textit{Ostriker} losses are $10^3-10^4$ orders of magnitude larger that the \textit{Type--I} losses around the crossing with the GW losses $r \sim 10-100 r_\isco$. At larger orbital separations the differences decrease. Unfortunately, the region $r \sim 10-100 r_\isco$ is crucial to understand when modeling an inspiral, so a difference this large invokes huge uncertainties. See section \ref{sec:typeIvsDF} for further discussion.

\subsubsection{Circular Inspiral}
\begin{figure}
    \centering
    \includegraphics[width=0.9\textwidth]{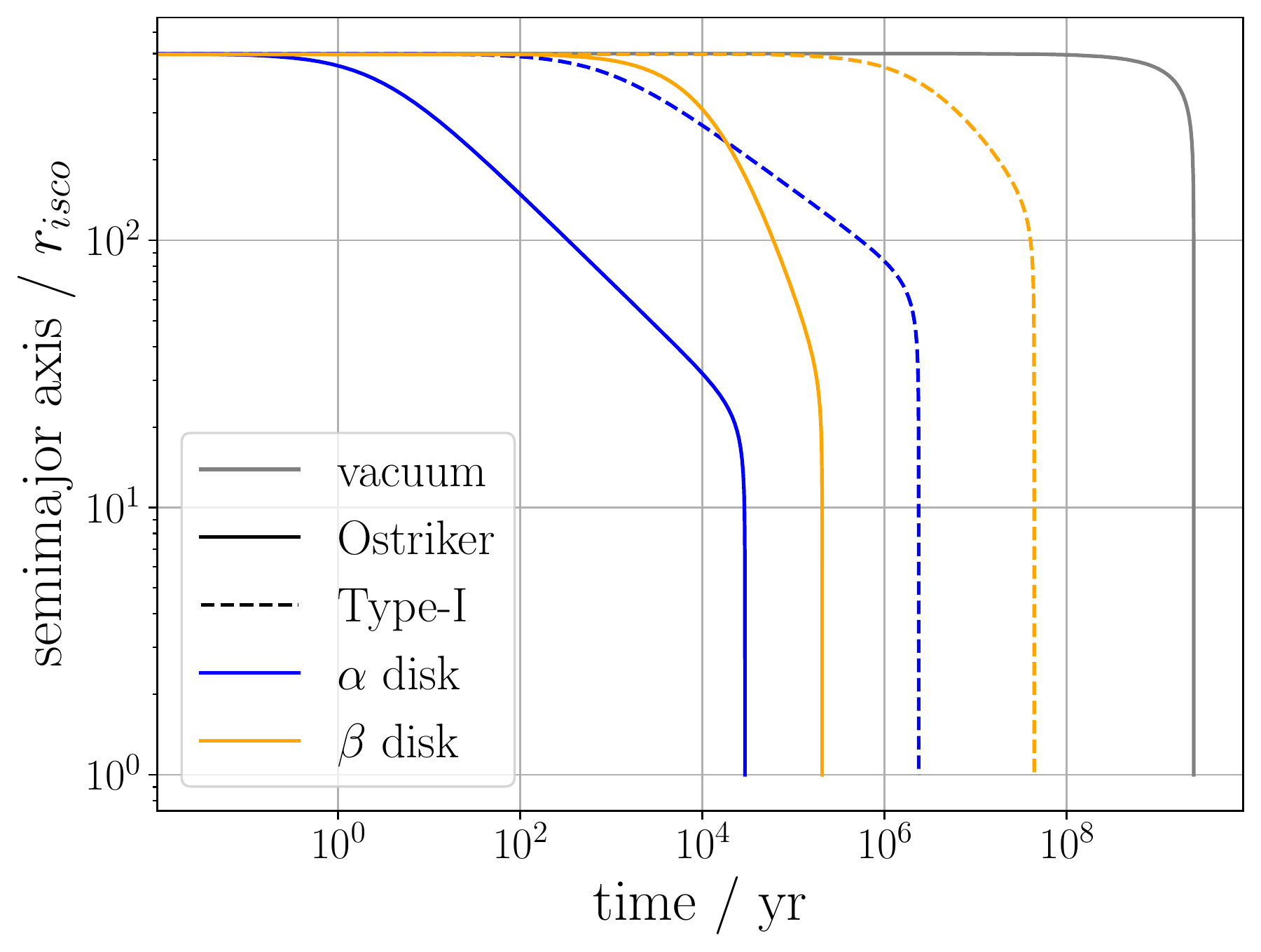}
    \includegraphics[width=0.9\textwidth]{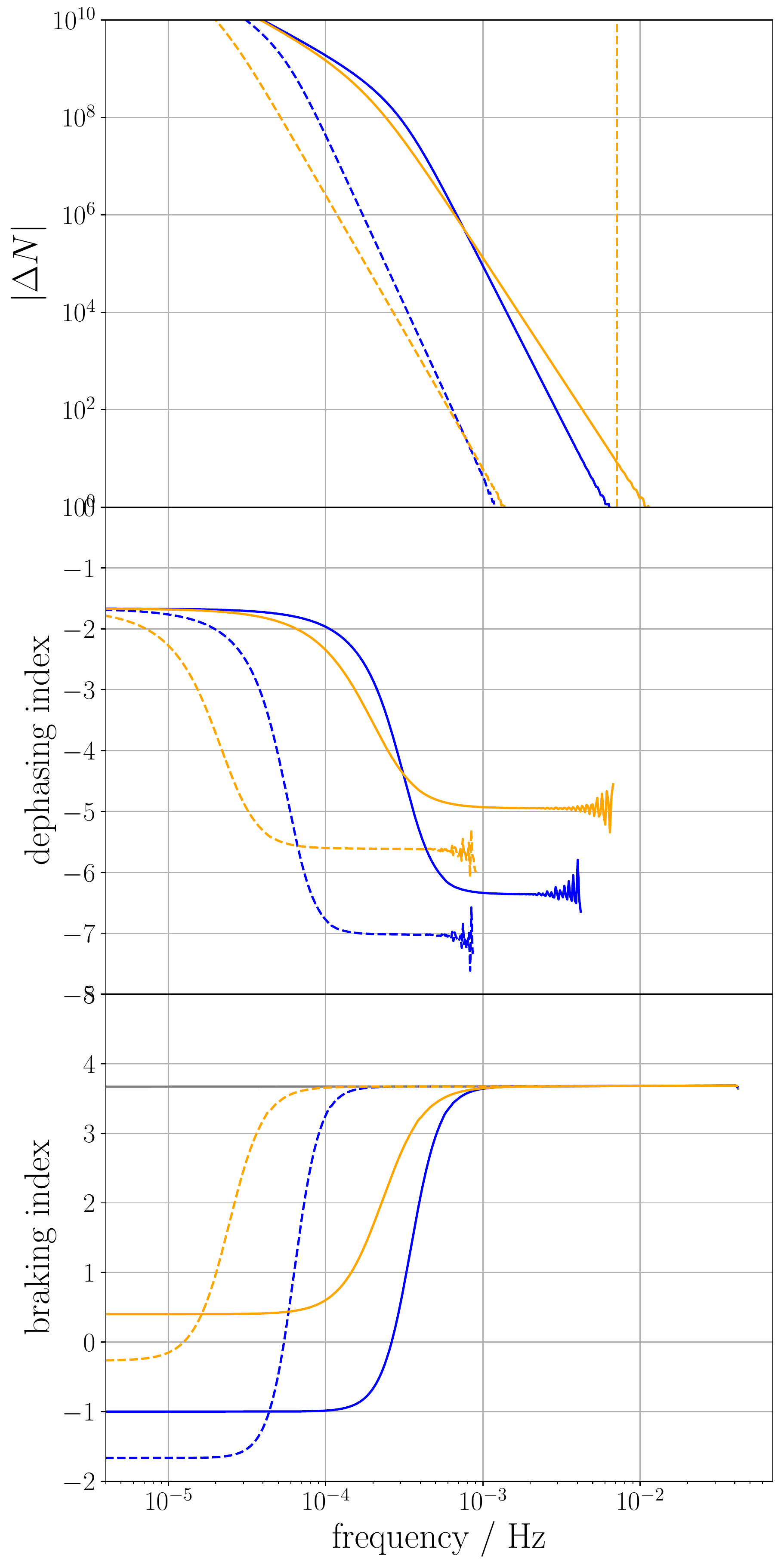}
    \caption{\textbf{Top}: The evolution of the semimajor axis $a$ for 5 different models, in a vacuum, in $\alpha$ or $\beta$ accretion disks, with \textit{Type--I} migration or \textit{Ostriker} dynamical friction interaction. \textbf{Bottom}: The dephasing, dephasing index, and braking index as a function of frequency during the inspiral. }
    \label{fig:ev_ba}
\end{figure}

To see the impact this has on an inspiral and the frequency evolution, see \figref{fig:ev_ba}. There, we model a circular inspiral with parameters $\{m_1, m_2, a_0\}= \{10^5 M_\odot, 1 M_\odot, 500 r_\isco \}$. In the top plots, the semimajor axis is plotted against time. The different magnitudes of the forces result in different timescales, with \textit{Ostriker} inspiraling orders of magnitude faster compared to the \textit{Type--I} model. Due to the higher density of the $\alpha$ disk at the most relevant separations, its inspiral is also faster compared to the one of $\beta$ disk. For the \textit{Ostriker} + $\alpha$ disk model, the semimajor axis, and with it the frequency evolution, significantly changes on the order of $\sim 10$ years, which would probably make the frequency evolution in this regime observable.\footnote{For our values of $a_0 = 500\, r_\isco \sim 3\cdot 10^3 m_1$, we are at the edge of the validity of \eqref{eq:alpha_disk}, so this effect might be exaggerated. Nevertheless, the trend seems to continue into the range of validity and warrants further inspection. }

In the second plot, which shows the dephasing amount, this trend is actually reversed. Here, the dephasing effects are stronger for the $\beta$ disk at late times, due to the crossing in relative impact seen in \figref{fig:avsb_T1vsOs} at small separations. For the \textit{Type--I} interaction, the dephasing is far below one in the last $5$ years of inspiral, which indicates that these effects will probably not be observable for these parameters. For the \textit{Ostriker} model, the dephasing ($\sim 100$) would also hardly be observable over a 5 year period.

Nevertheless, if the effects were to be observable, the dephasing index, shown in the third plot, would clearly distinguish between the forces. The lines are shown until $\Delta N < 10$ and they clearly converge to the value given by \eqref{eq:deph_index} for the different models. Complementary, the braking index, shown in the fourth plot, initially starts out at the value given by \eqref{eq:brak_approx} for the different models and converge to the value given for the GW emission loss.

The two plots of the braking and dephasing index clearly reflect the different regimes that the inspiral is subject to. Early in the evolution -- for low frequencies -- the braking index is constant with the expectation given by \eqref{eq:brak_approx}. Then, as the object inspirals and GW losses become important, the braking index moves in between these values and approaches the $n_b=11/3$ value at later times. At the same time, the approximation used to derive \eqref{eq:deph_index} becomes accurate, and the lines start to converge to the appropriate values. The two indices are clearly complementary observational probes.

Comparing the two CO accretion disk interaction models, while these results are certainly not sufficiently realistic, there stark contrast allows to make some inferences. If the \textit{Ostriker} description is close to reality, the inspirals will most likely be much faster and possibly have observationally relevant effects. On the other hand, for the \textit{Type--I} model, the inspirals will be more difficult to observe. 

Nevertheless, due to the nature of either forces as seen in \figref{fig:avsb_T1vsOs}, their effects would probably be more observable at larger separations.

\subsubsection{Eccentric Inspiral}
\begin{figure}
    \centering
    \includegraphics[width=0.77\textwidth]{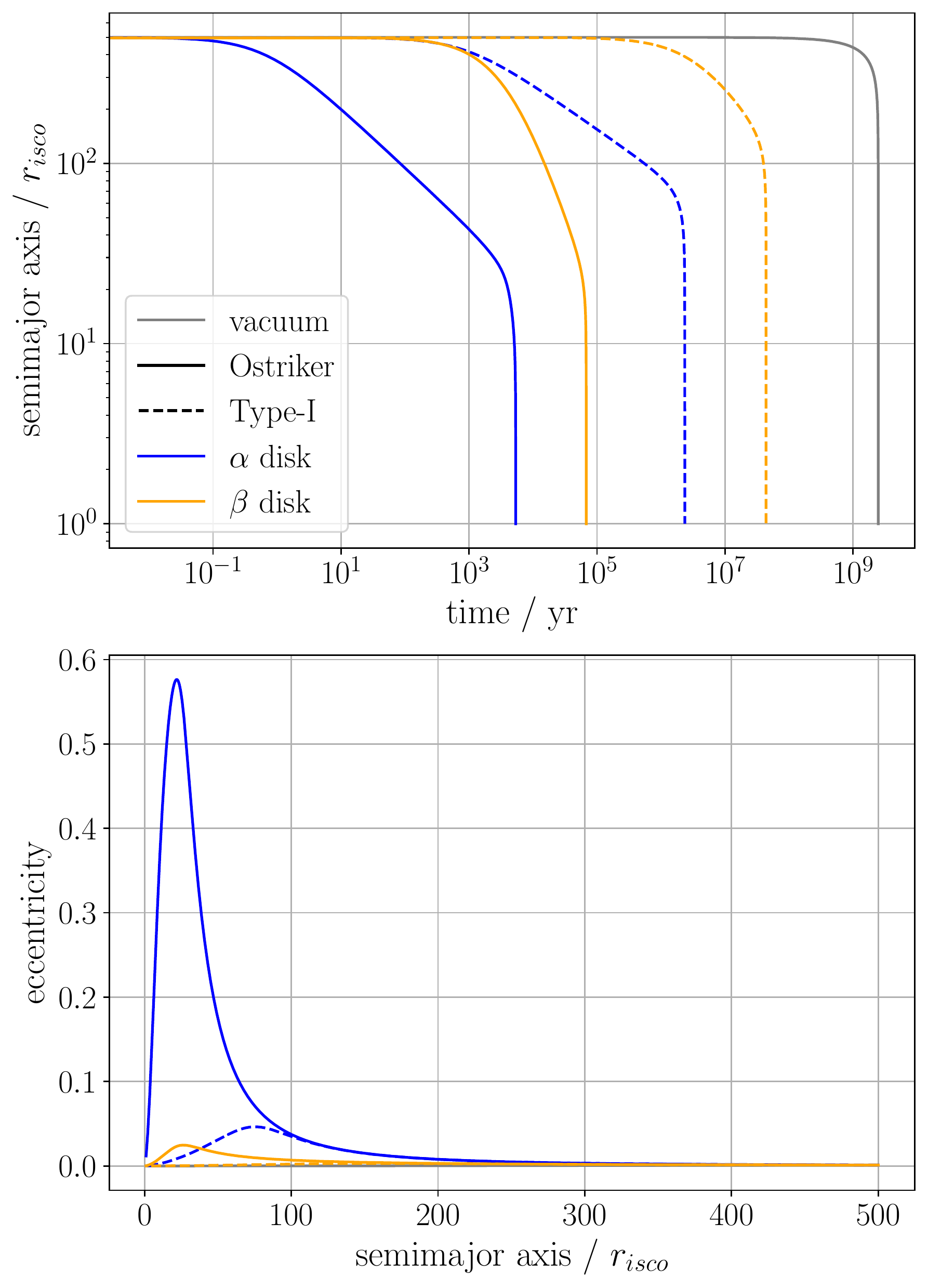}
    \includegraphics[width=0.77\textwidth]{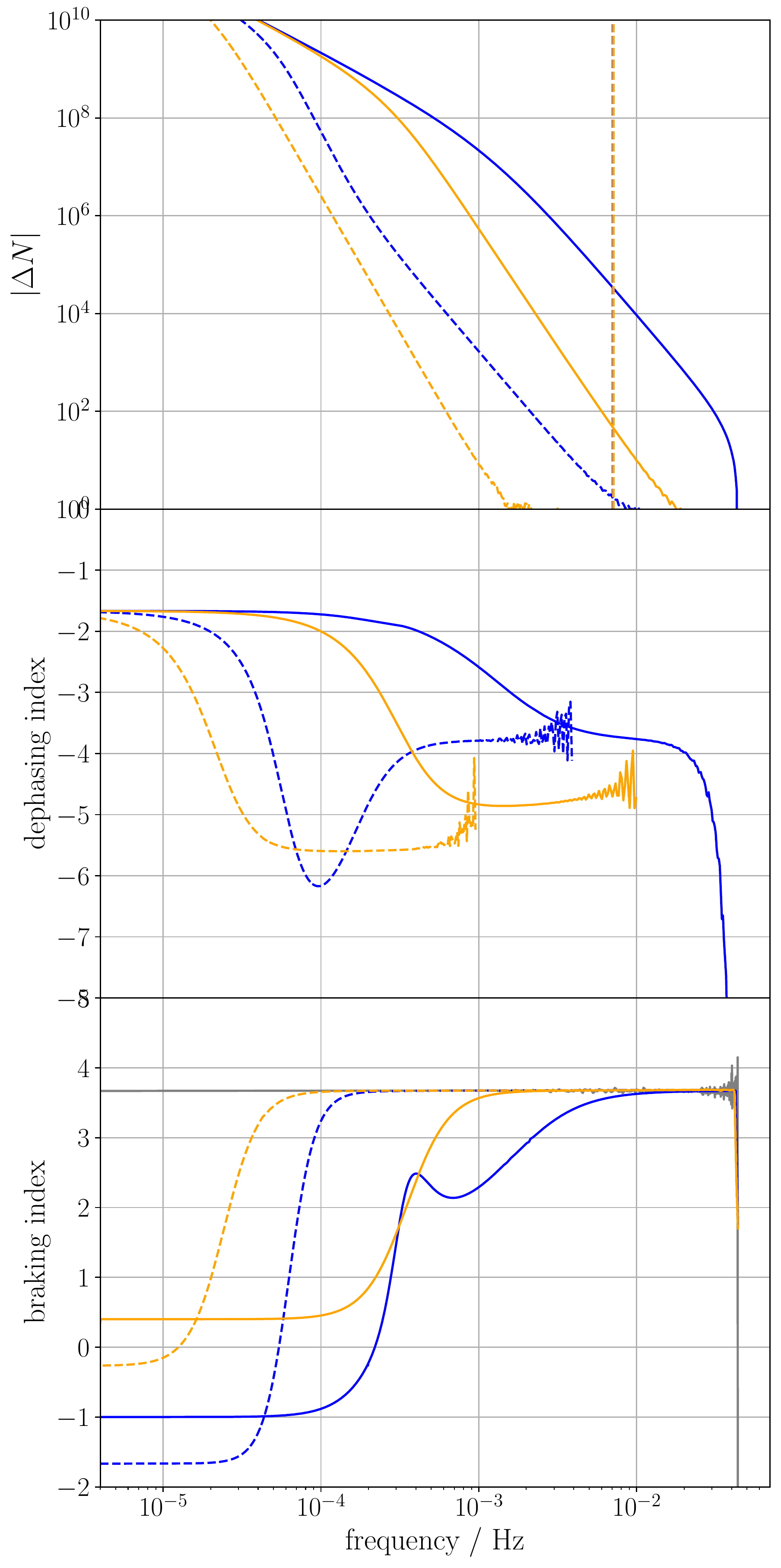}
    \caption{\textbf{Top}: The evolution of the semimajor axis and eccentricity for 5 different models, in a vacuum, in $\alpha$ or $\beta$ accretion disks, with \textit{Type--I} migration or \textit{Ostriker} dynamical friction interaction. \textbf{Bottom}: The dephasing, dephasing index, and braking index as a function of frequency during the inspiral. }
    \label{fig:ev_ba_ecc}
\end{figure}

If we allow for some small initial eccentricity $e_0 = 0.001$, we see a very different behavior. The eccentricity evolution is plotted in the second plot of \figref{fig:ev_ba_ecc}, and the temporal evolution is from right to left, from large semimajor axis to $r_\isco$.

First, the \textit{Type--I} migration model is only valid for circular orbits and breaks down for eccentricities $> 0.001$. The small eccentrification seen in the \textit{Type--I} model is most likely an artefact of the model being extended too simply to eccentric orbits and not physical \cite{Tanaka_2004}. However, the eccentricity evolution of the \textit{Ostriker} model is of interest. In both $\alpha$ and $\beta$ disks, the eccentricity increases, as dynamical friction tends to do. Of note are the different scales of the plots, the eccentrification is much stronger for the $\alpha$ disk. This can be understood with the condition derived in \cite{Becker:2021ivq}. For $F\sim r^\gamma v^\delta$, the eccentrification is proportional to $\dv{e}{t} \sim (1 - \delta + \gamma)$. For dynamical friction $\delta=-2$, for the $\beta$ disk $\gamma =-3/5$, while for the $\alpha$ disk $\gamma=3/2$. So the density distribution acts as an eccentrification moderator in the $\beta$ disk case, while the $\alpha$ disk density distribution enhances the eccentrification effects.

The strong eccentrification for the $\alpha$ disk also has an effect on the dephasing. Higher eccentricity means stronger GW emission loss, so eccentricity increases the inspiral rate, and therefore the dephasing. The dephasing at the $5$ year line increases by a factor of $10^2-10^4$ in the two models. Also, the dephasing index converges to a different value, as \eqref{eq:deph_index} was derived for a circular GW loss dominated inspiral. The different value is now due to the eccentricity increasing the inspiral rate. To tease out the accretion disk effects, one would need to expand \eqref{eq:deph_index} to eccentric GW loss inspirals. 

Similarly, the breaking index is affected by the eccentricity. This is due to the $\dv{e}{a}$ term in \eqref{eq:brak_approx}, whose behavior causes the small spike seen the last plot in \figref{fig:ev_ba_ecc}. Here, the braking index approximation would allow a measurement of $\gamma, \delta$ and therefore the profile of the disk ($\alpha$ disk with $\rho_\b \sim r^{3/2}$) and nature of the interaction (dynamical friction with $\delta = -2$).

We can conclude that by modeling the eccentric behavior it allows us to detect the environmental effect through the larger dephasing at late times, and additionally extract accurate information about the environmental effect via the braking index at early times. 

\FloatBarrier
\subsection{Accretion Disk + DM spike}

In the following section, we add a DM spike into the picture. For the spike we take the parameters $\{\rho_6, \alpha_\sp\} = \{1.3 \cdot 10^{17} M_\odot/\text{pc}^3, 7/3 \}$.

\begin{figure}
    \centering
    \includegraphics[width=0.95\textwidth]{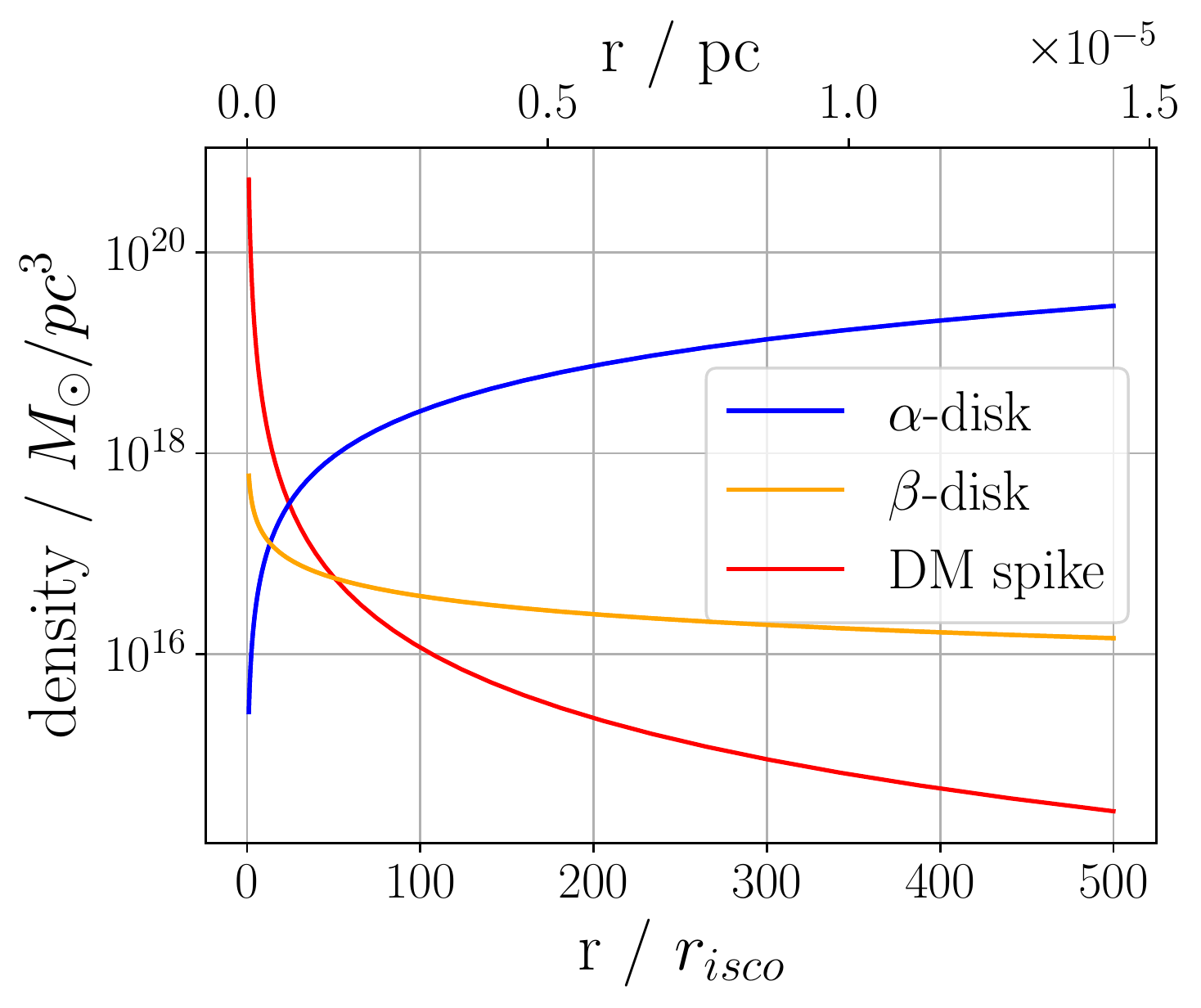}
    \includegraphics[width=0.95\textwidth]{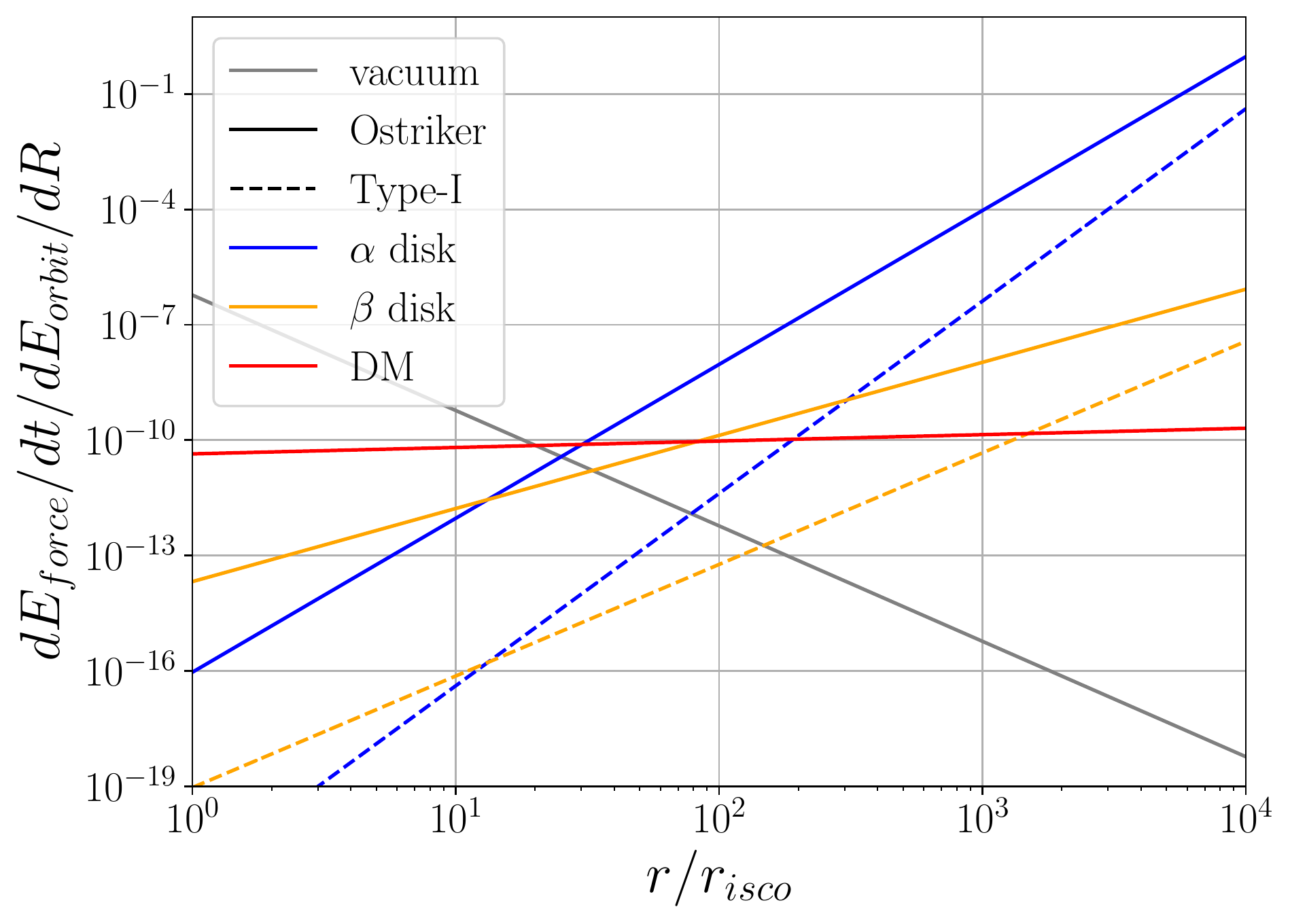}
    \caption{\textbf{Top}: The density $\rho_\b$ of the two accretion disk models $\alpha$ and $\beta$ and the DM spike $\rho_\dm$ around an IMBH with $m_1=10^5 M_\odot$. \textbf{Bottom}: The relative impact of the four dissipative forces: GW emission loss, DM dynamical friction, and \textit{Type-I} and \textit{Ostriker} for the two different disk models for a circular orbit at the given radius.}
    \label{fig:avsb_T1vsOs_dm}
\end{figure}

First, we add the DM distribution and relative impact to the previous comparative plot in \figref{fig:avsb_T1vsOs_dm}. It can be seen that the DM impact is on par with the \textit{Ostriker} model in the regime $r\sim 10-10^2 r_\isco$, but does not rise like the accretion disk effects. At smaller separations, it is subdominant to the GW loss, but stronger than the accretion disk effects.

To reduce the number of models, we focus on two combinations: The strongest effects are expected for \textit{Ostriker} + $\alpha$ disk, while the weakest effects are with \textit{Type--I} + $\beta$ disk. To get an idea of the possible relative impacts of DM spike vs accretion disks, we want to compare these combinations.

\FloatBarrier

\begin{figure}
    \centering
    \includegraphics[width=0.9\textwidth]{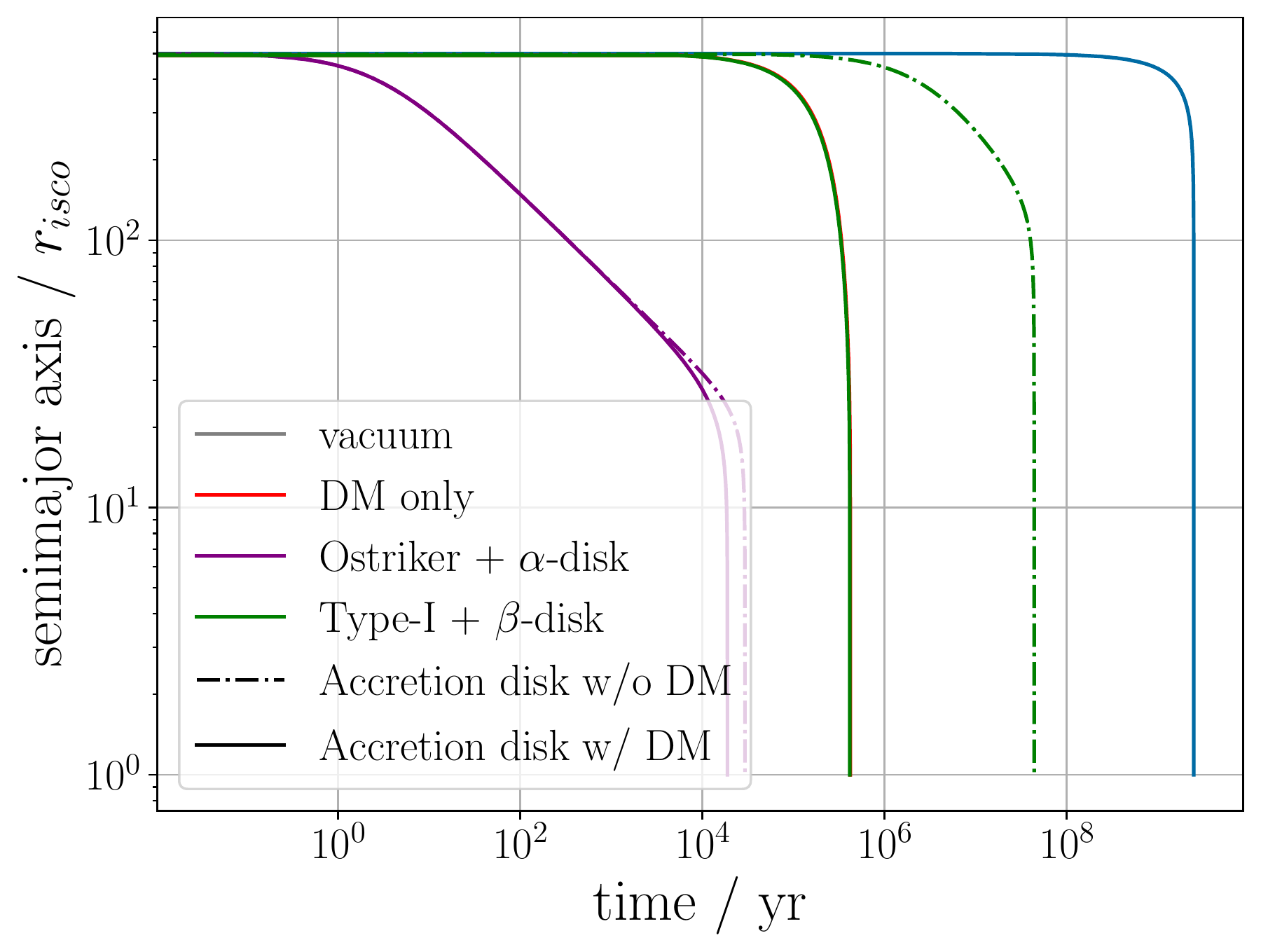}
    \includegraphics[width=0.9\textwidth]{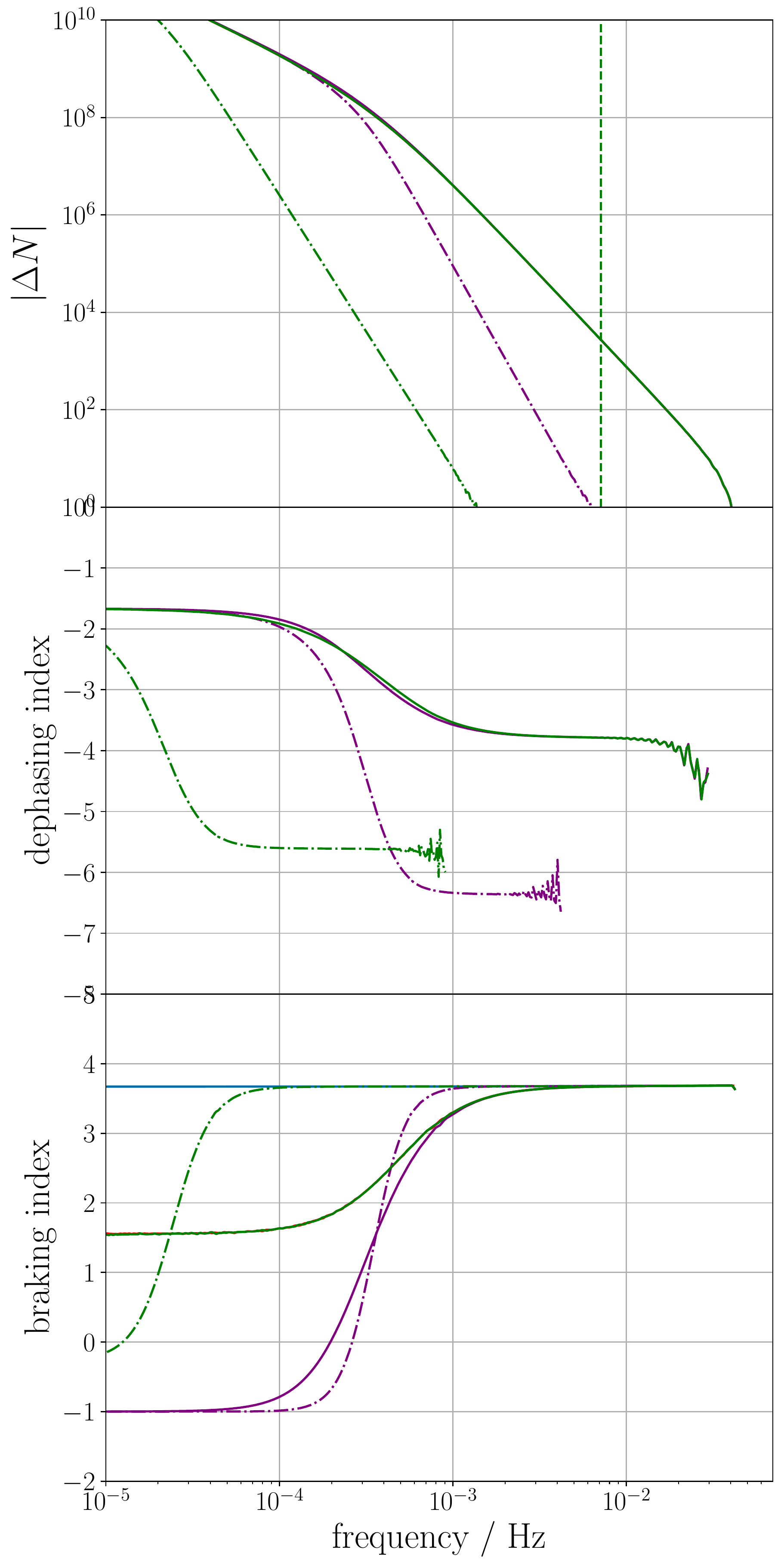}
    \caption{\textbf{Top}: The evolution of the semimajor axis $a$ for 6 different models, in a vacuum, with a DM spike, and $\alpha$ disk+\textit{Ostriker} and $\beta$ disk + \textit{Type--I}interaction with and without a DM spike. \textbf{Bottom}: The dephasing, dephasing index, and braking index as a function of frequency during the inspiral.}
    \label{fig:ev_ex}
\end{figure}


\subsubsection{Circular inspiral}

To this end, the evolution for a circular system with $m_1, m_2, a_0 = \{10^5 M_\odot, 1 M_\odot, 500 r_\isco \}$ is plotted in \figref{fig:ev_ex}. The results seem to be as anticipated. The fastest inspiral is for \textit{Ostriker} + $\alpha$ disk + DM spike, while the addition of the DM spike barely matters for the inspiral time, as \textit{Ostriker} + $\alpha$ disk is the dominant dissipative force initially. This can be seen in the braking index, where the evolution is dominated by its value. The dephasing on the other hand is clearly dominated by the DM spike. This can be understood by looking at \figref{fig:avsb_T1vsOs_dm}. The dephasing is accumulated where GW emission loss dominates, which is also where the relative impact of the DM spike is much stronger than the accretion disk effects.

The slowest inspiral (ignoring the vacuum case) is for \textit{Type--I} + $\beta$ disk. The addition of the DM spike clearly dominates this model. Still, at early times, the braking index would dominate the DM effects as inferred from \figref{fig:avsb_T1vsOs_dm}, but not for the range of radii seen here.

Overall, even though there are large modeling uncertainties, this implies that accretion disk and DM spike effects have different regimes of dominance, and could therefore be distinguished in an actual observation. DM spike effects are significant at small separations, while at large separations, accretion disk effects dominate. This is reflected in the braking and dephasing index.

\subsubsection{Eccentric inspirals}
\begin{figure*}
    \centering
    \includegraphics[width=0.85\textwidth]{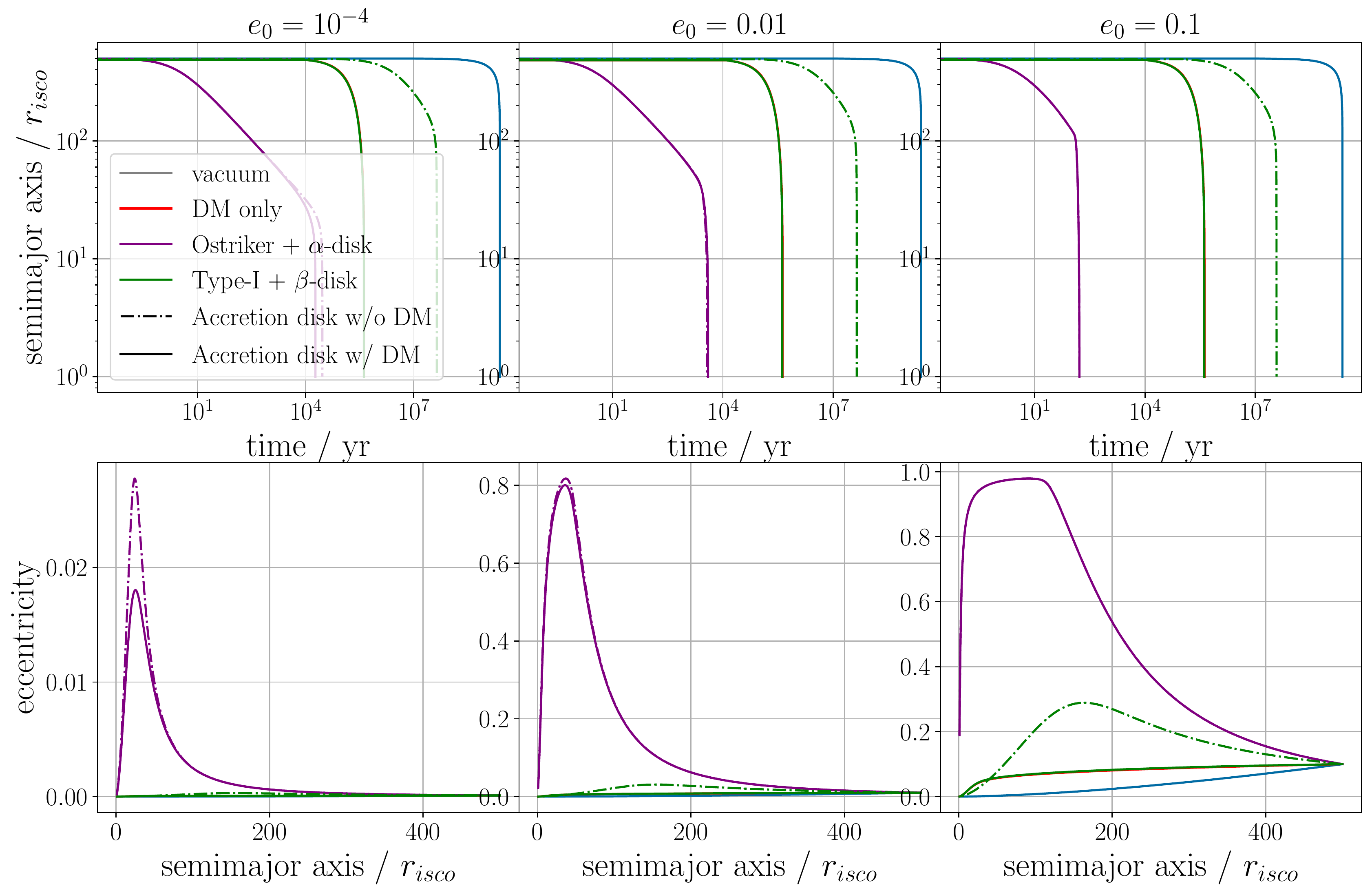}
    \includegraphics[width=0.85\textwidth]{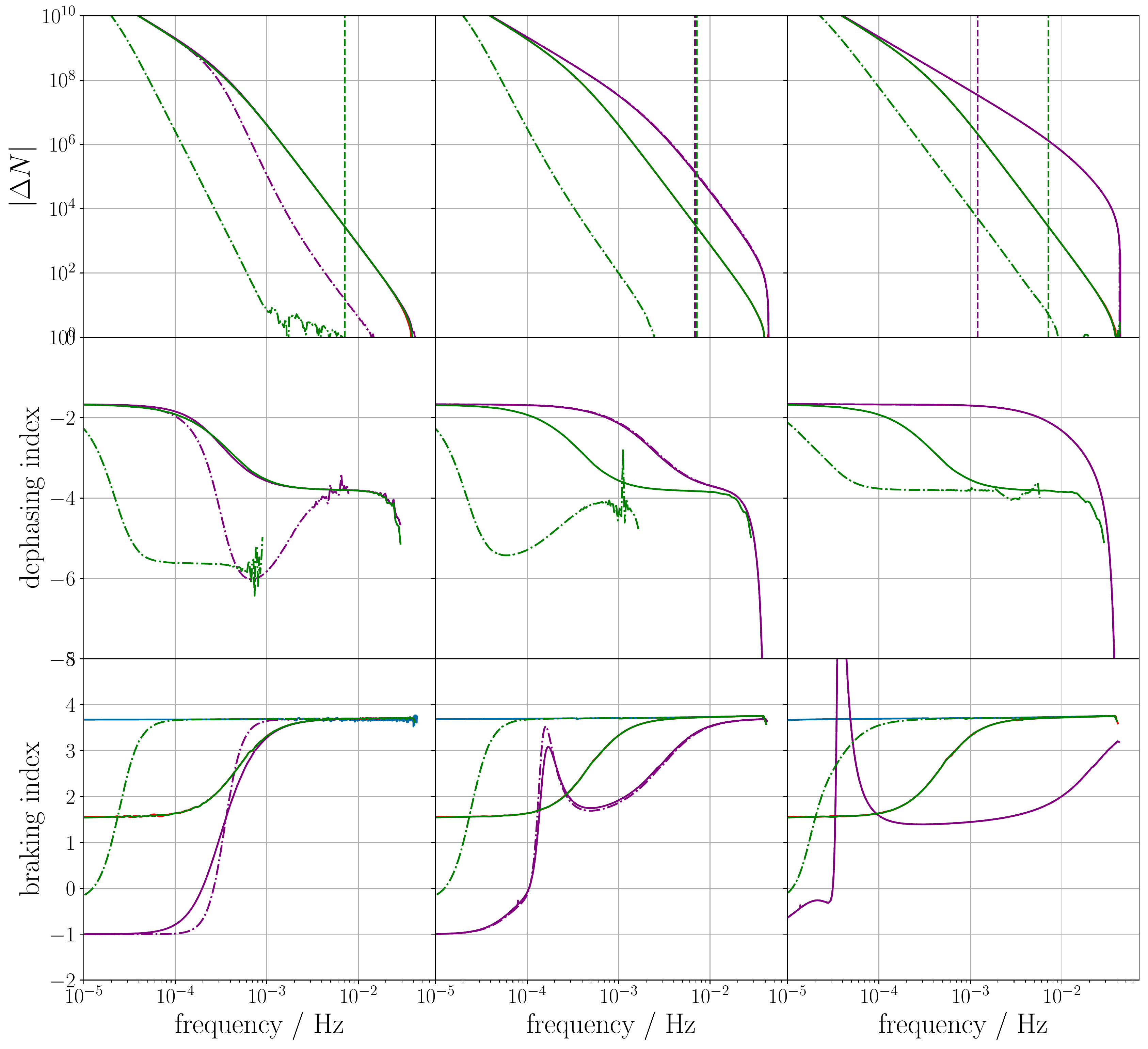}
    \caption{\textbf{Top}: The evolution of the semimajor axis and eccentricity for 6 different models, in a vacuum, with a DM spike, and $\alpha$ disk+\textit{Ostriker} and $\beta$ disk + \textit{Type--I} interaction with and without a DM spike for different initial eccentricities $e_0 = \{10^{-4}, 0.1, 0.6  \}$ \textbf{Bottom}: The dephasing, dephasing index, and braking index as a function of frequency during the inspiral.}
    \label{fig:ev_e0}
\end{figure*}

We now want to look at the orbital evolution allowing for eccentric orbits. To this end, we look at different initial eccentricities with $e_0 = \{10^{-4}, 10^{-2}, 10^{-1}\}$. The same caveats as described previously apply to the \textit{Type--I} migration model, so we will not comment on this model. The results are shown in \figref{fig:ev_e0}.

What can be seen again are the strong eccentrification effects of the \textit{Ostriker} + $\alpha$ disk model. For small initial eccentricity $e_0=10^{-4}$ the moderating effects of the DM spike as explored in \cite{Becker:2021ivq} can be seen. For higher initial eccentricity, the influence of DM is too weak initially, so the eccentricity (almost) saturates $e\to 1$. The higher eccentricity also means stronger GW emission, which speeds up the inspiral, causing very large dephasings for the \textit{Ostriker} + $\alpha$ disk model. There is competition between two effects described previously: DM causes circularization while also generally causing higher dephasing, whereas the \textit{Ostriker} + $\alpha$ disk model causes eccentrification and more dephasing through this. The flip can be seen in the comparison between the dephasing plots of $e_0=10^{-4}$ and $e_0 = 10^{-2}$.

What can be observed is the breakdown of the differentiation power of the dephasing index. Where eccentricity dominates, the models cannot be distinguished through the dephasing index. We would need better modeling of the circularization effects of the GW emission to pick out the forces involved. 

Overall, this reinforces the idea that the inspirals might be very eccentric if the \textit{Ostriker} + $\alpha$ disk model is accurate. The inspirals would be much faster and possibly within the lifetime of a spaceborne GW observation mission.

The initial eccentricity of the secondary most likely depends on the origin, whether it formed in-situ in the accretion disk, or whether it was captured \cite{Derdzinski:2022ltb}. The results here suggest that these different origins could be distinguished through the strongly different observational signatures.

\subsubsection{Varying Central Mass}
We now want to compare different central masses of the massive black hole to compare the impact in different regimes, therefore we vary $m_1 = \{10^3, 10^4, 10^5\}M_\odot$. According to \eqref{eq:alpha_disk} and \ref{eq:beta_disk}, the disk profiles change along with $m_1$. To allow for a fair comparison we vary the spike density $\rho_6 = \{5\cdot 10^{15}, 2.5\cdot 10^{16}, 1.3\cdot 10^{17} \}M_\odot/$pc$^3$ along with $m_1$.\footnote{Numerically employing the procedure described in section II.B of \cite{Eda:2014kra}, we obtain a geometric dependence of $\rho_6$ on the black hole mass $m_1$ such that $\frac{\Delta \log \rho_6}{\Delta \log m_1} \approx 0.7$. Therefore, increasing the black hole mass $m_1$ by a factor of $\sim 10$ increases $\rho_6$ by a factor of $\sim 5$. This allows us to generalize the parameters taken from \cite{Coogan:2021uqv}. } 

Unfortunately, the models we employ here begin to break down for larger mass ratios $q > 10^{-5}$. For example, halo feedback processes actually become important for the dynamical friction with the DM spike\cite{Kavanagh:2020cfn, Coogan:2021uqv}, which we do not model here. Also, \textit{Type--I} migration requires a smaller mass ratio, see sec. \ref{sec:typeIvsDF} for a discussion. Nevertheless, we still believe there is value in these plots as we will discuss in the following.

\begin{figure*}
    \centering
    \includegraphics[width=\textwidth]{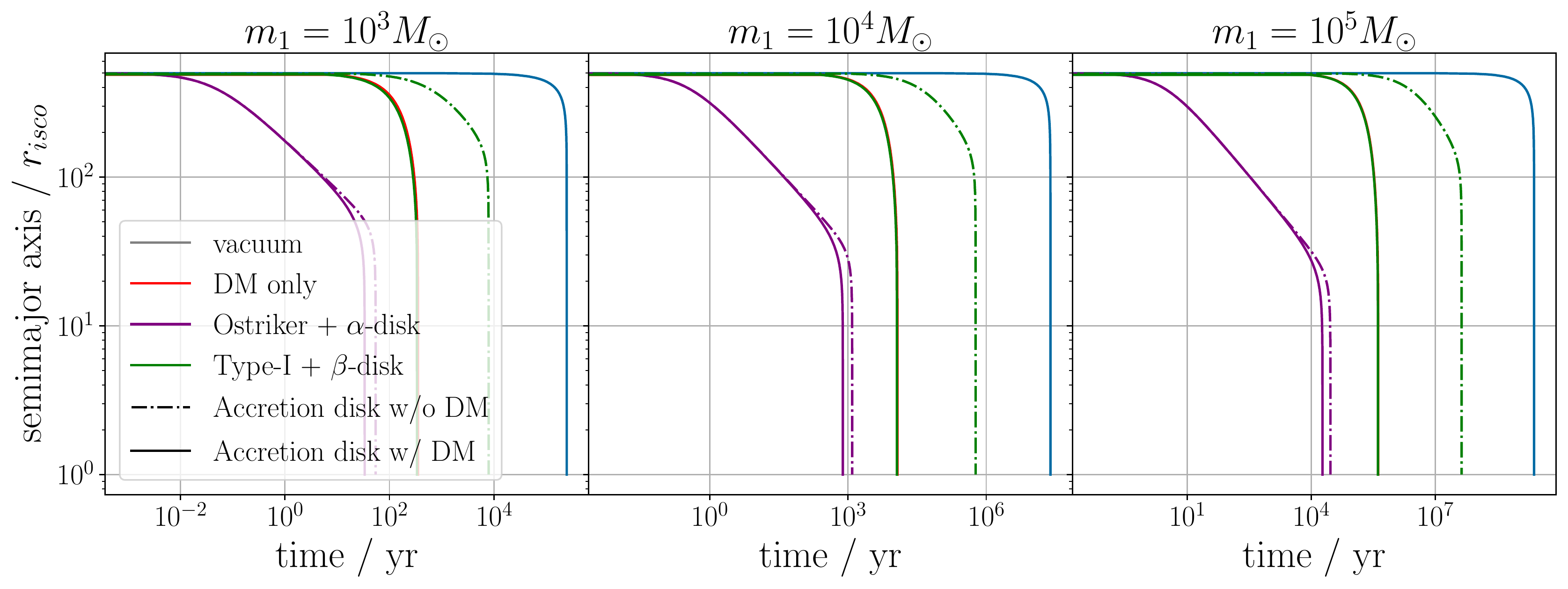}
    \includegraphics[width=\textwidth]{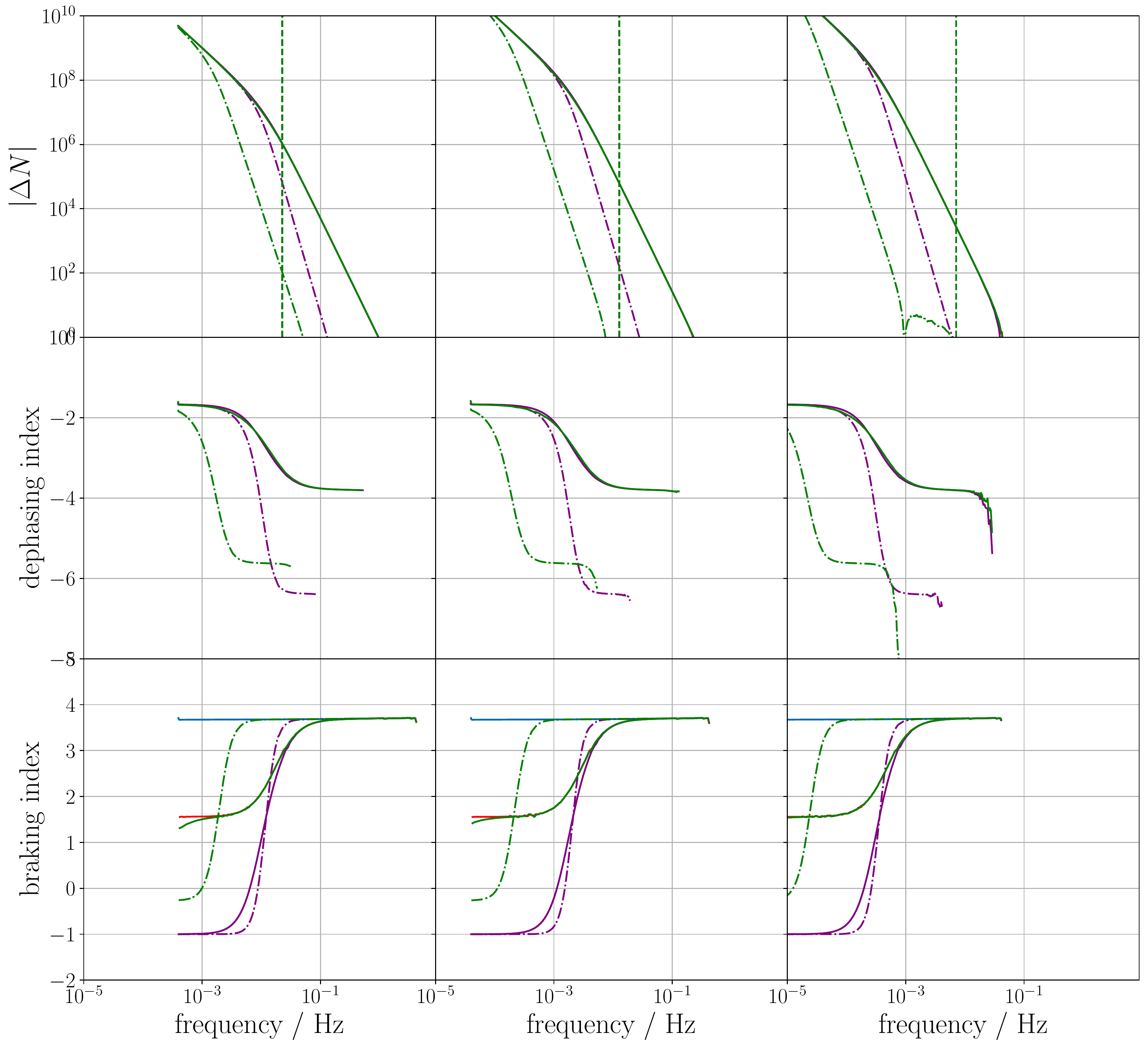}
    \caption{\textbf{Top}: The evolution of the semimajor axis and eccentricity for 6 different models, in a vacuum, with a DM spike, and $\alpha$ disk+\textit{Ostriker} and $\beta$ disk + \textit{Type--I} interaction with and without a DM spike for different IMBH masses $m_1 = \{10^3, 10^4, 10^5  \} M_\odot$ \textbf{Bottom}: The dephasing, dephasing index, and braking index as a function of frequency during the inspiral.}
    \label{fig:ev_m1}
\end{figure*}

Our results are plotted in \figref{fig:ev_m1}. It can be seen that the relative strength between the DM spike and the accretion disk effects does not change for central mass. The most prominent difference is the time of inspiral and the frequency range. The effects of the models are qualitatively the same, just shifted to the new frequency region and on a faster timescale. This implies that the detectability of these effects increases for smaller $m_1$ for the limited lifetime of a spaceborne GW observatory mission. We can see that the dephasing effects increase by many orders of magnitude for the last $5$ years of the inspiral. 

So even though these models break down, this is the region where a naive extension predicts much stronger observable effects. Therefore, we see a strong motivation to model this region of mass ratios and to better understand the forces involved. We leave this for future work.

\section{Discussion \label{sec:discussion}}

\subsection{Type--I migration vs Dynamical Friction in the Accretion Disk \label{sec:typeIvsDF}}
In this subsection we discuss the applicability of the models we employ for the CO -- accretion disk interaction.

The \textit{Type--I} torque model originates from protoplanetary disk models, where the mass ratio is sufficiently small, $q < 10^{-4}$. In this regime, linear perturbation theory describes the protoplanetary system very well\cite{1980ApJ...241..425G}, and is supported by numerical simulations \cite{2002ApJ...565.1257T}. For larger mass ratios, or larger COs, that have a size comparable to the disk, a gap can form in the accretion disk, reducing the torque experienced by the secondary. This is dubbed \textit{Type--II} migration.

Since the physics primarily depends on the mass ratio, an argument for the applicability of \textit{Type--I} can be made for IMRIs, where $q \ll 1$.  While the complete picture is more complicated, simulations have shown that this is a decent approximation to an order of magnitude for IMRIs \cite{Derdzinski:2018qzv, Derdzinski:2020wlw}. This model has been studied in the context of E/IMRIs in several previous publications\cite{Yunes:2011ws, Kocsis:2011dr, Barausse:2014tra,  Speri:2022upm, Cole:2022fir}. 
Additionally, \cite{Kocsis:2011dr} has argued that -- in AGN -- both migration types can appear during an inspiral. Further out in the orbit, the tidal field of the secondary would dominate over the central SMBH, and clear a gap. At some crossover radius, the tidal influence diminishes so that the secondary cannot clear a gap anymore, and the inspiral is dominated by \textit{Type--I}. See \cite{Derdzinski:2020wlw} for a discussion of how the torque can change during an inspiral in IMRIs.

The other model that is commonly applied to these systems, is that of \textit{dynamical friction}, which we have previously referred to as \textit{Ostriker}. Here, the secondary creates a wake behind itself in the orbit and the resulting gravitational interaction slows it down \cite{1987gady.book}. Ref. \cite{Canto:2012bg} derives that this model is relevant for a hypersonic secondary ($v^2 \ll c_s^2$) in thick disk models, in which it is completely embedded inside the disk. The model is typically used for eccentric orbits \cite{Sanchez-Salcedo:2019wjx, Sanchez-Salcedo:2020pae, 10.1093/mnras/stac1294} but is not strictly necessary \cite{2015ApJ...811...54G, Kim:2007zb}. Simulations show that dynamical friction is an accurate approximation for a point like secondary, e.g. a black hole, as long as it is completely embedded in the disk and the mass ratio is small enough $q \lesssim 10^{-5}$ \cite{Sanchez-Salcedo:2020pae}. Even though we have used a thin disk description here, the secondary might be sufficiently small to be completely embedded, i.e. the Roche radius is smaller than the disk $r_\textit{roche} \ll H$.
For circular prograde orbits (where the secondary would be co-rotating with the disk), this model breaks down, as the relative velocity with a differentially rotating disk would be $0$. For this reason, we focused on retrograde orbits relative to the disk. These models can be relevant in capture scenarios, where the secondary was not produced inside the disk, but captured from the intergalactic medium \cite{Sanchez-Salcedo:2020pae}.

Comparing the two models, \cite{2015ApJ...811...54G} points out that they are closely related. The forces are proportional by a factor of $F_\ostriker/F_\typeI \propto (r/H)$, which comes down to the \textit{differential torque} \cite{1986Icar...67..164W}. This is the difference between the inner and outer torque on the secondary in orbit. \textit{Type--I} migration modeling is sensitive to this torque, while \textit{dynamical friction} in the local ballistic approximation is not. This is analogous to including tidal forces acting on the secondary \cite{2015ApJ...811...54G}. This can be seen in \figref{fig:avsb_T1vsOs}, where the difference is a scaling in $r$ (since $H\sim$const here). The difference becomes more pronounced for small $r$, which is unfortunately where the effects are most pronounced in an observable inspiral. 
Ultimately, whether these differential torques are relevant depends on multiple factors, such as the density gradient of the disk, the gravitational size (i.e. Roche radius) of the secondary, its direction and speed relative to the disk. While there is numerical support by simulations for both models, each simulation has a limited range of applicability and caveats.

The conclusion in this discussion can only be that more study is needed to assess which effects are most relevant at these scales.

\subsection{Comparison between different environmental effects}
Understanding the different environmental effects in E/IMRIs is of great importance to maximize the science yield of future space based GW detectors \cite{Barausse:2014tra, Zwick:2022dih}. 

The two possible environmental effects we have explored and compared are (baryonic) accretion discs and DM spikes. For simplicity, we also have focused on a single DM spike model. 

The results seem to indicate that DM spikes and accretion disk effects dominate at different times in the inspiral. At early times and larger separations, accretion disk effects dominate, while at late times and small separations, DM spike effects are stronger. These regimes are reflected in the different braking and dephasing indices. Another factor is the circularization effect of DM spikes, explored in \cite{Becker:2021ivq}, which competes with the eccentrification effect of the \textit{Ostriker} model. This can influence the inspiral even if DM is subdominant.
From these initial considerations we would carefully conclude that the environmental effects can be distinguished and -- from the perspective of trying to detect DM -- the accretion disk effects are sufficiently different and not superdominant as to allow a detection. If both accretion disks and DM spikes are common around IMBH, some IMRIs should reflect the accretion disk effects while they are at large distance, while others could reflect the DM spike influences when they have smaller separations.

One aspect left unexplored is that of halo feedback. As the secondary loses energy and angular momentum to the DM spike, it can actually significantly deplete the spike locally \cite{Kavanagh:2020cfn}. This effect is more prominent for larger mass ratios $q \gtrsim 10^{-4}$. We chose our parameters ($m_1 \sim 10^5M_\odot$) such that for most of the systems explored in the previous section, halo feedback would be negligible. The results where $m_1 < 10^5M_\odot$ should be taken with a grain of salt. As explained previously, the same relative impact between the forces appeared for different $m_1$, which would break down with the inclusion of halo feedback.
But we can make some inferences what including halo feedback processes would mean for the model. For example, if DM dynamical friction would be the dominant force, the reduction in the DM density due to the halo feedback might increase the relevance of the accretion disk effects. On the other hand, if the accretion disk effects significantly dominate and the halo feedback timescale is larger than that of the accretion disk effects, halo feedback might be less relevant. Overall, the DM spike influence would be harder to observe. We leave these considerations for future studies.

It is important to keep in mind that we only study linear combinations of environmental effects. In reality, all these environmental effects would be interacting with each other and possibly deviate from their simplistic descriptions. Exploring these interactions will have to be done with numerical simulations. 

\subsection{Observational Signatures}
While there are large modeling uncertainties in these systems, fortunately the observational signatures differ between these models. For late time inspirals, where the GW emission dominates, the dephasing is a tool to observe the environmental effects\cite{Barausse:2014tra}. The amount of dephasing and the speed at which it is accumulated depends on the dissipative force. If the derivative of the dephasing accumulated can be measured, the dissipative force can in principle be identified by the dephasing index, as given by \eqref{eq:deph_index}.

For larger separations of the secondary, an inspiral might still be too far off to be observable within the mission lifetime, considering the timescales involved. At the same time, for larger separations environmental effects would dominate over the GW emission loss, dictating the frequency evolution. If the evolution of the frequency is observable, i.e. the second derivative $\ddot{\mathcal{F}}$ is measurable, the braking index can be used to differentiate between environmental effects as well. According to \eqref{eq:brak_approx}, different dissipative forces can result in different frequency evolutions. Depending on the physical expectations, the forces involved can be inferred. If at the same time the eccentricity evolution can be measured, the specific force can be pinned down.

The observability of the second derivative was estimated in \cite{Robson:2018ifk} and the braking index for inspiraling binaries in a common envelope in \cite{Renzo:2021aho}. These results indicate that it might be observable for some systems, but most of the observable systems would be stationary. Nevertheless, if the dissipative force is strong enough, for example as in the \textit{Ostriker} model in \figref{fig:ev_ba}, the frequency evolution could be sped up for the effect to be observable during the observational period of LISA. We leave the detectability of the braking index for different dissipative forces for future study.

Just like in equal mass binary systems, modeling the eccentricity evolution is important for E/IMRIs \cite{Bhat:2022amc}. The eccentricity evolution can have large effects on the frequency evolution and the dephasing. When measured, it can also hint at the environmental effects at play, as hinted by \eqref{eq:brak_approx}. 

All of these observational tools, the braking index, the amount of dephasing and dephasing index, and the eccentricity evolution, are complementary. The braking index is valuable for large separations when the environmental effects dominate. The dephasing and dephasing index are important late in the inspiral, when GW emission loss dominates. The eccentricity evolution complements both of these tools and can help to pin down the environmental effect(s) involved. 

Of course, we employed a simplistic description of the environmental effects with a force $F\propto r^{\gamma} v^{\delta}$. More general forces could be dependent on other features, such as spin, tidal deformability, etc. If the environmental effects are time-dependent, these indices would vary over time, losing their descriptive value. But even for more complex environmental effects, such as the halo feedback \cite{Kavanagh:2020cfn, Coogan:2021uqv}, a new equilibrium can emerge and the indices become distinct, as can be seen from the phase parameterization they develop \cite{Coogan:2021uqv}. Eventually, relativistic and post-Newtonian effects will have to be analyzed in this framework. We will look at generalizations of these indices in a future study. 

During the finalization of this publication, \cite{Cole:2022fir} have published their results. They compared the inspiral waveforms of DM spikes, accretion disks and scalar clouds and performed a Bayesian analysis to see if these environments can be distinguished. They find that these environments can in fact be easily distinguished by their features. Our analytic approximations might shed a light on these numerical results. 

It could be useful to map out different environmental effects, their braking and dephasing index and eccentrification/circularization effects. These tools  would allow an abstraction of the environmental effects and focus on the specific impact on the frequency evolution. This could generalize waveform generation and put a handle on the large parameter spaces that are incurred when looking at several environmental effects. 

As LISA expects to see a plethora of overlapping signals from different source classes, the collaboration's strategy is to attempt a global fit with all the possible source class parameters. Depending on the behavior of these classes and their predominant physical effects, their braking index might be an identifier of a class, and could therefore help distinguish between the different classes. 


\section{Conclusions \label{sec:concl}} 
In this paper, we studied the effect of CO -- accretion disk interactions on IMRIs. We compared two models -- \textit{Type--I} migration and \textit{Ostriker} dynamical friction interaction with the accretion disk \cite{Ostriker:1998fa}-- and explored the effects they have on the evolution of the semimajor axis and eccentricity in an inspiral. Then, we added a dark matter spike and compared the effect with the two baryonic accretion models. 

\begin{itemize}
    \item The relative impact of \textit{Type--I} migration and \textit{Ostriker} dynamical friction differs by several orders of magnitude. In a  \textit{Type--I} migration scenario, we expect no eccentricities and negligible dephasing effects. In a retrograde dynamical friction scenario we expect large eccentricities, large dephasing and a domination of the frequency evolution for typical IMRI systems.
    \item In comparison to the DM, the interactions have different regimes of dominance, DM is more dominant for small separations and accretion disk effects at larger separations.  
    \item We are able to differentiate between the models individually and in combination, due to different amount of dephasing, a difference in braking and dephasing index, and a different impact on the eccentricity evolution.
\end{itemize}
Which of the two baryonic accretion disk interactions models is more accurate remains to be studied. We also leave the inclusion of halo feedback effects to further study.

We would like to emphasize the usefulness of the study of the braking and dephasing index and eccentrification effects in distinguishing environmental effects and leave a systematic study of environmental effects and their properties for future studies.



\begin{acknowledgments}
We thank Andrea Derdzinski, Yury Levin and the anonymous referee for insightful discussions and helpful comments on the manuscript. 
\end{acknowledgments}

\appendix

\section{Derivation of braking and dephasing index \label{sec:app_deriv}}
Assuming we have a dissipative force of the form $F(r,v) = F_0 r^{\gamma}v^{\delta}$, then plugging it into \eqref{eq:avgdEdT}, (\ref{eq:da_dt}) gives
\begin{align}
    \dot{a} =& \avg{\dv{E_F}{t}}/\pdv{E_\orb}{a} \\
    =&  -\frac{2F_0}{\mu} a^{k_1} (1-e^2)^{k_1 + 1/2} m^{(\delta-2)/2} \nonumber \\
    & \cross \underbrace{\int_0^{2\pi} (1 + e \cos\phi)^{-(2+\gamma)}(1 + 2e\cos\phi + e^2)^{(\delta+1)/2}}_{\approx 1 + k_2 e^2}
\end{align}
with $k_1 = 2 + \gamma - \frac{\delta + 1}{2}$ and $k_2 = (3 + \gamma^2 +\gamma(3-2\delta) - 2 \delta + \delta^2)/4$. The approximation of the integral is valid to third order in $e$. This equation here rectifies a wrong positive sign in comparison to \cite{Becker:2021ivq}.

Taking the time derivative gives
\begin{align}
    \ddot{a} ={}& \dot{a} \left( k_1 \frac{\dot{a}}{a} - 2e\dot{e}\frac{k_1-1/2}{1- e^2} + 2e\dot{e} \frac{k_2}{1+k_2e^2}  \right) \nonumber \\
    = {}&\frac{\dot{a}^2}{a} \left(k_1 + 2ae \dv{e}{a} \left( \frac{1/2 - k_1}{1- e^2} + \frac{k_2}{1+k_2 e^2} \right)  \right)
\end{align}
which immediately results in \eqref{eq:brak_approx}.

If we now assume that (circular) GW emission is the dominant force with $\dot{\F}_\vac \propto \F_\vac^{11/3}$ \cite{Cutler:1994ys}, and our dissipative force is a small addition to that with $\dot{\F}_F$, we can model this as the frequency evolution being the sum of both contributions $\dot{\F}_\text{tot} = \dot{\F}_\vac (1 + \varepsilon)$. A calculation of $\varepsilon$ gives 
\begin{align}
    \varepsilon ={} &  \frac{\dot{\F}_F}{\dot{\F}_\vac} \nonumber \\
    \propto {}& \, \frac{1}{\dot{\F}_\vac} \left( a^{-1/2} \avg{\dv{E_F}{t}} \right)\nonumber \\
    \propto {}& \, \frac{1}{\F_\vac^{11/3}} \left( a^{-1/2 + k_1 - 2} \right) \nonumber \\
    \propto {}& \, \F_\vac^{-11/3 - 2/3 (- 5/2 -2 k_1/3) } = \F_\vac^{-2 -2 k_1/3}
\end{align}
where we have assumed that $e=0$ and, in the last step, approximate $\F_\text{tot} \approx \F_\vac$ such that $\F_\vac \propto a^{-3/2}$. 


\bibliographystyle{apsrev4-1}
\bibliography{biblio}{}
\end{document}